# Symmetry-Breaking of Turbulent Flow in Porous Media Composed of Periodically Arranged Solid Obstacles

**Vishal Srikanth[1], Ching–Wei Huang[1], Timothy S. Su[1], and Andrey V. Kuznetsov[1]†**

[1]Department of Mechanical and Aerospace Engineering, North Carolina State University, Raleigh, NC 27695, USA

The focus of this paper is a numerical simulation study of the flow dynamics in a periodic porous medium to analyze the physics of a symmetry-breaking phenomenon, which causes a deviation in the direction of the macroscale flow from that of the applied pressure gradient. The phenomenon is prominent in the range of porosity from 0.43 to 0.72 for circular solid obstacles. It is the result of the flow instabilities formed when the surface forces on the solid obstacles compete with the inertial force of the fluid flow in the turbulent regime. We report the origin and mechanism of the symmetry-breaking phenomenon in periodic porous media. Large Eddy Simulation (LES) is used to simulate turbulent flow in a homogeneous porous medium consisting of a periodic, square lattice arrangement of cylindrical solid obstacles. Direct Numerical Simulation (DNS) is used to simulate the transient stages during symmetry breakdown and also to validate the LES method. Quantitative and qualitative observations are made from the following approaches – (1) macroscale momentum budget, (2) 2D & 3D flow visualization. The phenomenon draws its roots from the amplification of a flow instability that emerges from the vortex shedding process. The symmetry-breaking phenomenon is a pitchfork bifurcation that can exhibit multiple modes depending on the local vortex shedding process. The phenomenon is observed to be sensitive to the porosity, solid obstacle shape, and the Reynolds number. It is a source of macroscale turbulence anisotropy in porous media for symmetric solid obstacle geometries. In the macroscale, the principal axis of the Reynolds stress tensor is not aligned with any of the geometric axes of symmetry, nor with the direction of flow. Thus, symmetry-breaking in porous media involves unresolved flow physics that should be taken into consideration while modeling flow inhomogeneity in the macroscale.

---

## 1. Introduction

### *1.1. Background*

Microscale turbulence in porous media possesses dual character with features of both classical internal and external turbulent flows. Unique flow phenomena are anticipated when the two features interact inside the porous medium. In this paper, we analyze a symmetry-breaking phenomenon in the turbulent flow field that results in both micro- and macro- scale flow deviation. We define symmetry-breaking as the deviation of the resultant flow field in a symmetrically posed problem. Therefore, the geometry and the flow conditions must share symmetric axes. The flow deviation originates in the microscale flow field, specifically the solid obstacle surface forces, before it is transferred to the macroscale. In this work, the microscale is defined as the range of length scales that are smaller than the pore size (consistent with Nield 2002). Using the Volume Average Theory (VAT) (Slattery 1967), we are reducing the dimension of the microscale flow properties to obtain macroscale flow variables. We present the origin and mechanism of symmetry-breaking by using numerical simulation data.

† Email address for correspondence: avkuznet@ncsu.edu

Previous research in porous media has aimed to provide a universal description of the flow for model development. Early attempts employed Reynolds Averaging (RA) to further reduce the order of macroscale turbulence models. For a detailed account of the development of macroscale turbulence models in porous media, see de Lemos (2012), Lage *et al.* (2007), Vafai (2015), Vafai *et al.* (2009). These models enabled the study of transport processes in physical systems as applied to canopy flows, pebble-bed nuclear reactors, heat exchangers, porous chemical reactors, crude oil extraction, to name a few applications (Jiang *et al.* 2001; Mujeebu *et al.* 2014; Wood *et al.* 2020). However, there are a lot of unanswered questions about the flow physics of turbulence in porous media.

Initial attempts to study the microscale flow in porous media used Reynolds Averaged Navier-Stokes (RANS) simulations (Kundu *et al.* 2014; Kuwahara & Nakayama 1998; Pedras & de Lemos 2003). However, the information that is extracted from the microscale RANS simulations is limited by the modeling error (Iacovides *et al.* 2014). To overcome this, the Large Eddy Simulation (LES) approach was used to study the microscale flow physics and develop LES-VAT models (Jouybari & Lundström 2019; Wood *et al.* 2020). Microscale LES has also been used to determine turbulence statistics for model closure (Kuwahara *et al.* 2006; Kuwata & Suga 2015a; Suga 2016). The use of LES is a leap forward in revealing the transport of large-scale turbulence (Zenklusen *et al.* 2014).

There are two key observations about turbulence in porous media that are relevant to this work. The first is the proposition of the Pore Scale Prevalence Hypothesis (PSPH) (Jin *et al.* 2015; Uth *et al.* 2016), which introduced the notion of a pore-scale turbulence mixing layer (Jin & Kuznetsov 2017). Spatiotemporal scale suppression was also confirmed by the DNS studies of He *et al.* (2018, 2019). The second is the contrast in turbulence anisotropy properties between the micro- and macro- scale introduced by the volume average operation (Chu *et al.* 2019). It is also noted that the turbulence anisotropy in the bulk of the flow diminishes with an increase in the Reynolds number. This finding is consistent across DNS studies (Chu *et al.* 2018; He *et al.* 2019) and PIV measurements for packed beds (Khayamyan *et al.* 2017; Nguyen *et al.* 2019; Patil & Liburdy 2015).

Turbulent flow statistics in porous media have been extensively studied by researchers, but there is a lack of understanding about the dynamics of turbulent structures, particularly microscale vortices. From PSPH, microscale vortices (micro-vortices) are the source of large-scale turbulent structures in porous media. They cause microscale flow field inhomogeneity (Linsong *et al.* 2018) and are responsible for the strong geometry-dependence that has been observed in porous media flows (Vafai & Kim 1995). Both attached and detached vortex systems are observed in porous media flows. In the turbulent flow regime in porous media, the von Karman instability is observed in the detached vortex system that is formed behind the solid obstacles (Kuwahara *et al.* 2006). Descriptions of von Karman vortex shedding for the flow around a single cylinder, its origin, and manifestation can be found in von Karman (1911), Boghosian & Cassel (2016), and Durgin & Karlsson (1971). Unlike in a classical flow around a cylinder, a von Karman vortex street cannot be formed in porous media (see section 3.1.2). The von Karman instability acts in a confined space and has significant repercussions for the flow in porous media.

Several instances of vortex-induced flow instabilities and bifurcations have been reported in porous media. Yang & Wang (2000) reported a bifurcation that occurs at the pore-scale in periodic porous media that also affects the macroscale flow field. It results in the possibility of either a symmetric or an asymmetric vortex rotation in the microscale beyond a critical value of inlet velocity. A Hopf bifurcation is also observed to occur in porous media (Agnaou *et al.* 2016; Zhang 2008). Beyond a critical value of Reynolds number, the flow in the porous medium begins to oscillate due to the unsteadiness that is present in the vortex system. Spectral analysis of the transverse velocity has revealed that the unsteady flow is characterized by distinct frequencies at lower Reynolds numbers, with a greater degree of disorder (more frequencies) appearing as the Reynolds number increases. A detailed description of the origin and the mechanism of these phenomena has not been reported yet.

The present work focuses on a symmetry-breaking bifurcation phenomenon, which causes macroscale flow to deviate from the principal axes of symmetry (referred to hereafter as deviatory flow). The phenomenon was first observed by Iacovides *et al.* (2013, 2014), and West *et al.* (2014) in the context of cross flow in in-line circular tube bank heat exchangers. The deviation from symmetry is said to increase when the tubes are placed closer together. The deviation is attributed to the flow's tendency to follow the path of least resistance. The phenomenon is further examined by Abed & Afgan (2017) who have also made use of in-line circular tube banks. Asymmetry in the location of the "separation shear layer" and the blockage influence of downstream tubes are believed to characterize this phenomenon. However, the origin of this phenomenon and the details of its occurrence remain unknown. It is important to understand the mechanics of this phenomenon due to the abundance of porous media with low porosity utilized throughout various applications (Nield & Bejan 2017).

In this paper, the dynamic flow development from symmetry to symmetry breakdown is investigated. It is shown in section (3.1.1) that symmetry-breaking is effectuated through the amplification of a microscale imbalance in the pressure distribution that arises from the von Karman instability. Several geometric criteria determine the occurrence of symmetry-breaking, which are explored in section (3.2), but the parameters that control it are limited to the momentum supply and the magnitude of confinement. The phenomenon suggests a strong dependence of the macroscale field variables on the interaction between the micro-vortices. In practice, macroscale amplification of microscale instability provides a platform for enhanced fluid mixing in porous media and a source for macroscale instabilities that are larger than the pore scale.

Modeling symmetry-breaking in the turbulent flow regime in flows with low porosity requires an understanding of the underlying flow physics since the phenomenon is strongly dependent on geometry and boundary conditions. The mechanism by which the microscale events are transferred to the macroscale variables, which would result in macroscale symmetry-breaking, is dependent on a range of parameters, some of which are identified in section (3.2). To elucidate the interplay between microscale and macroscale variables, a comprehensive momentum budget is also presented.

### *1.2. An overview of the regimes of turbulent flow in porous media*

The microscale flow patterns in the porous medium are determined by the geometry. This is supported by evidence from the present work, and also Chu *et al.* (2018), Iacovides *et al.* (2014), Suga *et al.* (2017), and Uth *et al.* (2016). In section (3.2), we show that the porosity of the porous medium is one of the fundamental parameters that influences the symmetry-breaking phenomenon. The macroscale flow characteristics are strongly influenced by the porosity as a result of the change in the microscale flow field. In this section, we summarize the influence of porosity on the microscale flow patterns for circular cylinder solid obstacles.

The porosity ($\varphi$) changes the volume of fluid in the porous medium, which explicitly influences the macroscale flow. For porous media with circular cylinder solid obstacles, we mark $\varphi = 0.8$ as the boundary between high and low porosity. At low porosity ($\varphi < 0.8$ in this work), there is a strong interaction between the flow around adjacent solid obstacles. Here, the term interaction refers to the impingement of micro-vortices on the neighboring solid obstacles and the influence of the solid obstacle surface forces. Both streamwise and transverse flow interaction are observed. The symmetry-breaking phenomenon reported in this paper occurs at low values of porosity where the properties of the flow enable the amplification of flow instabilities. At high porosity, the flow behind a solid obstacle has a weaker interaction with the neighboring solid obstacles in the porous medium when compared to the flow in low porosity. The interaction can be strong in the streamwise direction, but it is virtually non-existent in the transverse direction.

The distinct regions of primary flow and secondary (vortex) flow that are observed (see Figure 1(a)) will change with the porosity. At high values of porosity ($\varphi > 0.8$), the core diameter of the micro-vortices scales with the diameter of the solid obstacle. At low values of porosity ($\varphi < 0.8$), the core diameter of the micro-vortices is limited by the size of the void space, leading to smaller micro-vortices than at high porosity. The change in the length scale of the largest flow structures with the porosity is more severe in the case of symmetry-breaking. It will influence the turbulence transport process in porous media flows.

## 2. Computational Details

### *2.1. Simulation Conditions*

The geometry of a porous medium is often intricate with significant spatial variation that can potentially alter the flow within. In this work, the porous medium is abstracted into a Homogeneous Generic Porous Matrix (GPM) that consists of solid obstacles populated in a simple square lattice. The diameter of the solid obstacle ($d$) and the pore size ($s$) are used to define the GPM. Setting aside the reduction in computational cost, the use of a homogeneous GPM facilitates a reduction in the number of variables in the problem. The idea is analogous to that of classical Homogeneous Isotropic Turbulence research. A Representative Elementary Volume (REV) of the GPM with periodic boundaries is used for simulations, which has the dimensions of $4s \times 4s \times 2s$. A cubic domain of size $s$ is adequate to represent the geometry of the GPM. However, the turbulent two-point correlations will de-correlate after a distance of $s$ (Jin *et al.* 2015; Uth *et al.* 2016). Accounting for the decorrelation width ($2s$) and boundary effects, a Turbulence REV (REV-T) size of $4s$ is chosen (see Figure 1(b)). Cylindrical solid obstacles are chosen to represent an anisotropic porous medium, similar to the geometry of heat exchangers. The sizing requirement for the REV-T is relaxed to $2s$ in the direction of cylinder axis due to a lack of geometric variation. It should be noted that several researchers have successfully utilized smaller domains for turbulence simulations in porous media (Iacovides *et al.* 2014; Kuwahara *et al.* 2006). The adequacy of the size of the REV-T has been analyzed and the results are presented in Appendix A.

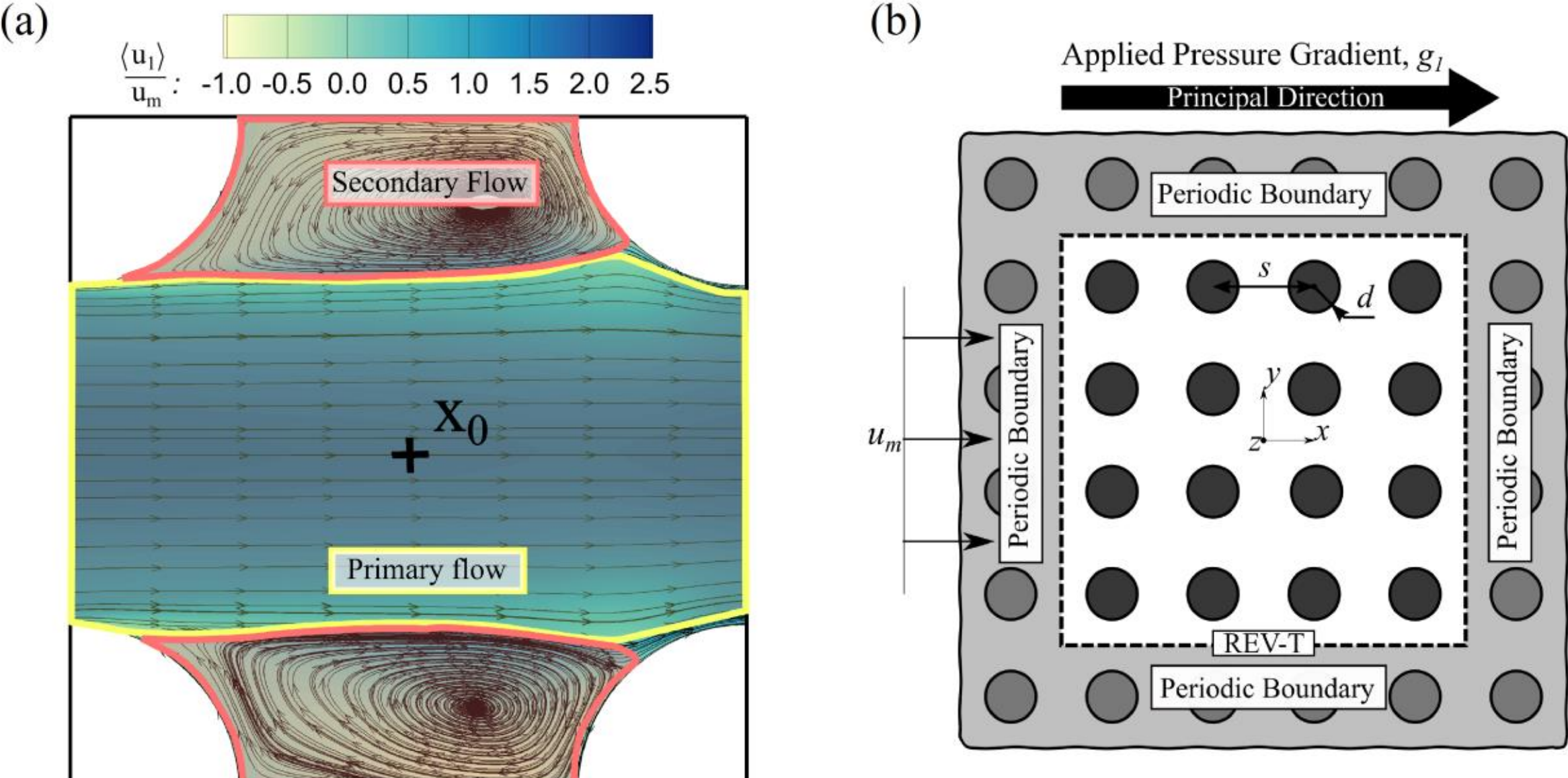

Figure 1: (a) Mean flow streamlines overlaid on contours of the time-averaged, microscale $x$-velocity $\langle u_1 \rangle$ at the midplane normalized by the time-averaged, macroscale $x$- velocity $u_m$. The primary (yellow) and secondary (pink) flow region boundaries are illustrated. (b) 2-D view of the GPM. The REV-T is used as the size of the computational domain, shown by dashed lines.

A Reynolds number of 1,000 is used in this work to study geometric parameter variation. The definition of the Reynolds number ($Re_p$) is provided in equation (2.1), where $u_m$ is the time-averaged, superficially averaged $x$- velocity, $d$ is the obstacle diameter, and $\nu$ is the kinematic viscosity of the fluid. Note that the operator $\langle - \rangle$ denotes Reynolds time-averaging. In literature, several definitions of the characteristic length scale of the porous medium are used (He *et al.* 2019; Li *et al.* 2020; Wood *et al.* 2020). The obstacle diameter is chosen to remain consistent with previous work (Jin *et al.* 2015). At $Re_p$ = 1,000, the flow through a porous medium is within the fully turbulent regime (Seguin *et al.* 1998a).

$$Re_p = \frac{u_m d}{\nu} \tag{2.1}$$

Both Large Eddy Simulations (LES) with a subgrid scale model and fine-grid Direct Numerical Simulations (DNS) without any turbulence model are used in this work. Both LES and DNS simulations are performed using the commercial CFD solver Ansys® Academic Research Fluent, Release 16.0. Unstructured grids constructed with a Block-Structured topology are used for the simulations. For LES, the grids are designed as per the guidelines established by Chapman (1979) and Choi & Moin (2012). The aspect ratio of the cells in the bulk of the computational domain is set equal to 1. The maximum cell size is of the order of the Taylor Microscale of the flow, which is estimated using the $Re_p$ as given in a book by Pope (2000). Setting the filter cut-off width equal to the Taylor microscale has been shown to give good agreement of classical Smagorinsky LES results with DNS for channel flows (Addad *et al.* 2008). Grid stretching with a cell growth ratio of 1.1 is used near the walls such that the maximum value of $y^+$ is less than one for the first grid point in the wall-normal direction. For DNS, the maximum grid size is set equal to the estimated Kolmogorov length scale. The Kolmogorov length scale is estimated using the Reynolds number according to the Kolmogorov hypothesis. The DNS grid is qualitatively identical to the LES grid with regard to its structure. The smallest scale of turbulence is expected to be less than the Kolmogorov scale estimate since the flow is wall-bounded. This compromise is accepted since the finest scales of turbulence are not of interest in this paper. The objective of the high-resolution DNS is to minimize the numerical error. Moser & Moin (1987) noted that a majority of turbulence dissipation happens at scales 15 times larger than the Kolmogorov scale for curved channel flows. The energy spectra had at least 100:1 reduction in the small scales when compared to the largest scale, which is confirmed in Appendix B. Details of the grid sizes used for the simulations are presented in Appendix B.

The Dynamic One Equation Turbulent Kinetic Energy (DOETKE) subgrid model (Kim & Menon 1997) is used for the LES cases. RANS turbulence models are a cheap alternative to LES provided their calibration is suitable for porous medium flows. However, Rodi (1997) showed that the performance of the $k-\varepsilon$ RANS models is inferior to LES for bluff body flows. The LES results show close agreement with experiments (Jin *et al.* 2016) while providing information about the dynamics of the large eddies. An additional transport equation for subgrid turbulent kinetic energy is solved in this work to model subgrid scales near the solid obstacle wall at higher Reynolds numbers. Davidson & Krajnovic (2000) demonstrated the effectiveness of the DOETKE model in predicting parameters associated with vortex shedding in the absence of adequate mesh resolution, despite the absence of the near wall turbulence streaks in the resolved flow field. Abba *et al.* (2003) brought out the shortcomings of the DOETKE model, which showed poor performance in predicting the near-wall turbulence dissipation. The DOETKE model assumes that the subgrid scales are isotropic in nature, which is reasonable if the subgrid scales are in the dissipative scales within the purview of the Kolmogorov hypothesis. Such an assumption will fail in the near-wall regions, which are dominated by small-scale anisotropic structures that are subject to a large dissipation rate. To account for this, smaller cells are used near the wall with the help of grid stretching.

Details of the numerical methods have been presented in the next section. All of the simulations that are presented in this work are three-dimensional. Flow statistics have been averaged over 100 flow cycles for a single solid obstacle. The simulations have been run on the North Carolina State University Linux Cluster. Suggestive computation times for LES is 30,000 CPU-Hours, and DNS is 100,000 CPU-Hours (1 CPU-Hour = Computation Time in Hours for a single CPU).

### *2.2. Numerical Methods*

The numerical methods and models used in this paper are described in the following sections (2.2.1) and (2.2.2). The grid convergence study for the numerical methods and the details of the grid size are presented in Appendix B.

#### *2.2.1. Large Eddy Simulation (LES)*

The filtered Navier-Stokes equations, written in equations (2.2)-(2.3) (the tilde notation denotes spatial filtering), are solved in conjunction with the DOETKE subgrid model using the Finite Volume Method (FVM). The computational grid in the FVM implicitly applies a box filter. The pressure term $\tilde{p}$ in equation (2.3) corresponds to a filtered periodic pressure (terminology adopted from the documentation of ANSYS Inc. 2016). In periodic flows, the pressure gradient term in the governing equations can be decomposed into a constant pressure gradient term $\rho g_i$ and the gradient of the periodic pressure $\partial\tilde{p}/\partial x_i$. The sum of the periodic pressure and the linearly varying pressure is the static pressure. A transport equation for the subgrid turbulence kinetic energy $k_{SGS}$ (equation (2.4)) is solved to estimate the subgrid velocity scale. The subgrid length scale $\Delta$ is set equal to the cube root of the cell volume. The subgrid turbulence eddy viscosity is estimated using equation (2.5). The model constants $C_k$ and $C_\varepsilon$ are determined dynamically according to Kim & Menon (1997). A test scale solution is constructed from the grid scale solution by using a top hat test filter with a width $\hat{\Delta}$ that is equal to twice the size of the grid filter width $\Delta$. The grid filter width $\Delta$ is defined as the cube root of the grid cell volume. The similarity between the SGS stress tensor $\tau_{ij}$ and the test Leonard stress tensor $L_{ij}$ is invoked to determine $C_k$ (equations 2.6-2.7). The value of $C_k$ is limited by $-\mu/\left(k_{SGS}^{1/2}\Delta\right)$. Similarity between the dissipation rate at the grid level $\varepsilon_{SGS}$ and the test level $\varepsilon_{test}$ is invoked to determine $C_\varepsilon$ (equations 2.8-2.9).

$$\frac{\partial \widetilde{u_j}}{\partial x_j} = 0 \tag{2.2}$$

$$\frac{\partial \rho \widetilde{u_i}}{\partial t} + \frac{\partial \rho \widetilde{u_i}\widetilde{u_j}}{\partial x_j} = -\frac{\partial \tilde{p}}{\partial x_i} + \frac{\partial}{\partial x_j}\left[(\mu + \mu_{T,SGS})\left(\frac{\partial \widetilde{u_i}}{\partial x_j} + \frac{\partial \widetilde{u_j}}{\partial x_i}\right)\right] + \rho g_i \tag{2.3}$$

$$\frac{\partial k_{SGS}}{\partial t} + \frac{\partial(\widetilde{u_j} k_{SGS})}{\partial x_j} = \left[C_k k_{SGS}^{1/2}\Delta\left(\frac{\partial \widetilde{u_i}}{\partial x_j} + \frac{\partial \widetilde{u_j}}{\partial x_i}\right)\right]\frac{\partial \widetilde{u_i}}{\partial x_j} - C_\varepsilon \frac{k_{SGS}^{3/2}}{\Delta} + \frac{\partial}{\partial x_j}\left(\mu_{T,SGS}\frac{\partial k_{SGS}}{\partial x_j}\right) \tag{2.4}$$

$$\mu_{T,SGS} = C_k k_{SGS}^{1/2}\Delta \tag{2.5}$$

$$\tau_{ij} = -2C_k {k_{SGS}}^{\frac{1}{2}}\Delta\widetilde{S_{ij}} + \frac{2}{3}\delta_{ij}k_{SGS};\; L_{ij} = -2C_k {k_{test}}^{\frac{1}{2}}\hat{\Delta}\widehat{\widetilde{S_{ij}}} + \frac{1}{3}\delta_{ij}L_{kk} \tag{2.6}$$

$$C_k = \frac{1}{2}\frac{L_{ij}\sigma_{ij}}{\sigma_{ij}\sigma_{ij}};\; \sigma_{ij} = -\hat{\Delta}{k_{test}}^{\frac{1}{2}}\widehat{\widetilde{S_{ij}}};\; k_{test} = \frac{1}{2}\left(\widehat{\widetilde{u_k}\widetilde{u_k}} - \widehat{\widetilde{u_k}}\widehat{\widetilde{u_k}}\right) \tag{2.7}$$

$$C_\varepsilon = \frac{\hat{B} - (\partial\widehat{\widetilde{u_i}}/\partial x_j)(\partial\widehat{\widetilde{u_i}}/\partial x_j)}{\left((\mu + \mu_{T,SGS})\hat{\Delta}\right)^{-1}{k_{test}}^{3/2}} \tag{2.8}$$

where

$$B = \left(\partial\widetilde{u_i}/\partial x_j\right)\left(\partial\widetilde{u_i}/\partial x_j\right) \tag{2.9}$$

The spatial derivatives are approximated using a bounded second-order central scheme (according to the work of Leonard 1991) for the convective terms and a second-order central scheme for the viscous terms. The location of the pressure variable is staggered such that it is stored at the centroid of the face of the cell. The governing equations are solved in a segregated manner using a pressure-implicit scheme with splitting of operators (PISO). A second-order implicit backward Euler method is used for time advancement. The momentum source term $g_i$ is used to specify a linear pressure drop to sustain flow in the periodic domain. The value of $g_i$ required to sustain a flow with a given Reynolds number is not known a priori. For incompressible flow, the specification of the mass flow rate is sufficient to maintain a constant Reynolds number (see equation (2.1) for the definition of the Reynolds number). Therefore, $g_i$ is determined iteratively during the pressure correction step from the difference between the desired and the current mass flow rate. The fluid material is chosen to be water since the solver uses the dimensional form of the governing equations. The results presented in the paper are non-dimensional.

### 2.2.2. *Direct Numerical Simulation (DNS)*

The Navier-Stokes equations written in equations (2.10)-(2.11) are solved using the same Finite Volume Method (FVM) as in the case of LES. The pressure term $p$ in equation (2.11) corresponds to a periodic pressure. In this paper, the term DNS refers to a numerical simulation of the Navier-Stokes equations without using a turbulence model. The grid resolution for DNS is finer that that used for LES, as given in section (2.1) and Appendix B. It is noted in appendix B that the small-scale eddies do not contribute any new information in our simulations, and that only the large-scale turbulent structures are of interest.

$$\frac{\partial u_j}{\partial x_j} = 0 \tag{2.10}$$

$$\frac{\partial \rho u_i}{\partial t} + \frac{\partial \rho u_i u_j}{\partial x_j} = -\frac{\partial p}{\partial x_i} + \frac{\partial}{\partial x_j}\left[\mu\left(\frac{\partial u_i}{\partial x_j} + \frac{\partial u_j}{\partial x_i}\right)\right] + \rho g_i \tag{2.11}$$

The magnitude of $g_i$ is maintained a constant in the DNS simulation, since it was determined in the LES simulation with similar operating parameters. If the resulting Reynolds number is different from that of the corresponding LES simulation, the value that follows from the DNS simulation is reported.

## 3. Results and Discussion

In this section, the symmetry-breaking phenomenon is analyzed in detail and presented as follows:

i. First, the development of symmetry-breaking is analyzed from a macroscale perspective. A macroscale momentum budget is computed and the components that are relevant to the symmetry-breaking process are identified. Spectral analyses of the macroscale forces are used to contrast the flow properties before and after symmetry-breaking.
ii. With the knowledge of the relevant macroscale flow physics, the microscale flow field distribution and the turbulent structures are visualized to understand the relationship between the turbulent vortex motion and the symmetry-breaking phenomenon. The characteristics of the turbulent structures are determined a priori to isolate the swirling turbulent structures, which have a more significant contribution to symmetry-breaking.
iii. The influence that symmetry-breaking has on macroscale turbulence transport is determined. The porosity and the Reynolds number are parametrized for this study. The influence of the shape of the solid obstacles in the GPM on symmetry-breaking is briefly discussed.

Microscale turbulence was defined earlier as the turbulent structures that have a length scale smaller than the pore scale. In this work, there is no scope for the presence of any macroscale turbulent structures. Therefore, macroscale turbulence is the volume-average of the microscale turbulent flow field.

### *3.1. The origin and the development of the symmetry-breaking phenomenon*

In this section, a detailed study of the origin of symmetry-breaking and the mechanism of its development are presented. The transient stages of the symmetry-breaking phenomenon are simulated by using DNS with the intention of simplifying the data analysis. Unlike DNS, the use of LES will introduce additional terms in the macroscale momentum budget that depend on the grid resolution. The high grid resolution in DNS is favorable for flow visualization, especially while extracting 3D coherent turbulent structures using the Q- criterion. In this work, the choice of LES model and numerical schemes introduces sources of artificial dissipation and numerical error. These limitations will influence the critical Reynolds number for the transition to turbulence and also to the subsequent deviatory flow. This makes DNS the more desirable method to simulate the dynamic process. Circular cylinder solid obstacles are used to represent a porous medium (see Figure 1(b)) with a porosity of 0.50. The Reynolds number is increased from 100 – 10,000 across flow regimes consisting of unique properties as shown in Table 1.

The flow properties in cases A1-A4 contribute to the understanding of how the phenomenon develops, as demonstrated in the next section.

| Case ID | Reynolds number ($Re_p$) | Flow Properties |
| --- | --- | --- |
| A1 | 100 | Laminar – 2D flow structures |
| A2 | 225 | Turbulent – Emergence of 3D features in the flow structures |
| A3 | 300 | Turbulent – Completely 3D flow structures |
| A4 | 489 | Turbulent – Deviatory flow |
| A5 | 1,000 | Turbulent – Deviatory flow |
| A6 | 3,000 | Turbulent – Deviatory flow |
| A7 | 10,000 | Turbulent – Deviatory flow |

Table 1: The DNS cases A1-A4 simulated to analyze the development of symmetry-breaking. The LES cases A5-A7 simulated to study Reynolds number dependence. The solid obstacles are circular cylinders and the porosity is 0.5 for all of these cases.

#### *3.1.1. Macroscale momentum budget*

The macroscale momentum budget is computed using the macroscale momentum conservation equation (de Lemos 2012). The macroscale momentum conservation equations are derived from the Navier-Stokes equations (2.11) by applying the VAT. The authors note that other forms of the conservation equation may be adopted to suit the modeling requirements, such as in Lasseux *et al.* (2019). The time dependence of the conservation equation is retained. Applying this to the transient flow through the periodic porous medium, the macroscale spatial derivatives are eliminated to result in the following:

$$\underbrace{\rho \frac{\partial}{\partial t}\left(\varphi \langle u_i \rangle^i\right)}_{inertial} = \underbrace{\rho\varphi g_i}_{applied} + \underbrace{\frac{\mu}{\Delta V}\int_{A_{interface}} n_j \, \partial_j u_i \, dS}_{viscous\ drag} - \underbrace{\frac{1}{\Delta V}\int_{A_{interface}} n_i p \, dS}_{pressure\ drag} \tag{3.1}$$

The operator $\langle - \rangle^i$ denotes an intrinsic volume average in the fluid domain. $\Delta V$ is the total volume of the periodic REV, $A_{interface}$ is the interfacial area and $n_i$ is the normal vector of $A_{interface}$. The value of $\Delta V$ for all the cases with a 4x4 GPM is 3.2 x $10^{-5}$ $m^3$. The value of $A_{interface}$ can be calculated as $50(1-\varphi)\Delta V$. The sum of the macroscale pressure drag and viscous drag in equation (3.1) (referred to hereafter as macroscale pressure and viscous forces) gives the total drag $R_i$. In this paper, the term drag is used to denote any force acting on the solid obstacle surface. In this porous medium system, an applied pressure gradient is used to sustain mass flow and all the forces that oppose it are called drag regardless of the convention. The authors note that there is a different naming convention followed in the aerodynamics community. At steady state, the Reynolds average of the drag $\langle R_i \rangle$ is balanced solely by the contribution of the applied pressure gradient $\langle \rho g_i \rangle$. The applied pressure gradient acts as a source of mechanical energy in the REV. The drag force on the solid obstacles will act as a sink of mechanical energy in the REV. The drag force in the *x*- direction needs to be compensated for by the applied pressure to sustain the flow. In the case of symmetric flow in the *x*- direction, there are no forces acting in the *y*- or *z*- directions after Reynolds averaging. In the case of deviatory flow, the balance of forces is different from the symmetric case. The conservation of mechanical energy needs to be satisfied in all the 3 directions, even though the flow deviates from the direction of applied pressure gradient. Note that there is no applied pressure gradient in the *y*-direction to counteract the flow.

The question arises as to how conservation of mechanical energy can be satisfied in the *y*-direction. The answer lies in the balance in the components of drag force themselves. The macroscale pressure and viscous forces have equal magnitude and act in opposite directions along the *y*- axis resulting in a net zero force (see Figure 2(a)). The lack of symmetry in the solid obstacle surface stress distribution results in a balance between a positive pressure force and a negative viscous force in the *y*- direction. This concept is elucidated in Appendix C. Conservation of mechanical energy can thus be satisfied for the flow in the *y*- direction in the absence of applied pressure gradient.

The symmetry-breaking phenomenon is considered a bifurcation because the deviatory flow solution is equally probable in both the positive and negative *y*- directions (in Figure 2(a), the positive deviation is illustrated). The polarity of the macroscale *y*- direction viscous and pressure forces will interchange for the two possible symmetry-breaking solutions. Several modes of deviatory flow are possible from the two solutions, which will be discussed at a later stage. Therefore, the interplay between the components in the *x*- and *y*- momentum budget are different. In the *x*- direction, the pressure drag and the applied pressure gradient are dominant. Consider the case A1 (Table 1) of laminar flow through the porous medium. The balance of *x*-direction forces for case A1 is illustrated in Figure 2(b). The forces in the *y*- and *z*- directions are 5 orders of magnitude less than those in the *x*- direction. For case A1, the microscale flow is strictly 2D and the macroscale flow can be considered 1D. Even at laminar Reynolds numbers, the macroscale pressure force is 4 times the magnitude of the macroscale viscous force. It will become apparent later on that the macroscale pressure force and the microscale pressure distribution dominate the flow properties for all of the simulations. In the *y*- direction, the components of the drag – pressure and viscous, are dominant. It is prudent to bear this in mind while analyzing the dynamic response of the fluid flow.

For packed beds, the limit of the laminar flow regime is at a pore-scale Reynolds number of 180 (Seguin *et al.* 1998a). The onset of the fully-turbulent regime in finite porous media is gradual and can occur in the range $180 < Re_p < 900$ (Seguin *et al.* 1998b). The transition regime may not exist for periodic porous media since the flow field is continuously perturbed and fed back into the REV. Transition and intermittency may be possible only through the local re-laminarization of the flow. However, transition effects are not encountered in any of the simulations in this paper. Turbulence has been observed in infinite porous media for Reynolds numbers as low as 478 for circular cylinder solid obstacles (Uth *et al.* 2016). In this work, the dynamic nature of the flow emerges at a Reynolds number of $Re_p = 225$ (case A2, Table 1). The macroscale momentum budget for case A2 is plotted against time in all the three directions in Figure 3. Note that the applied pressure gradient is maintained at a constant value for this case.

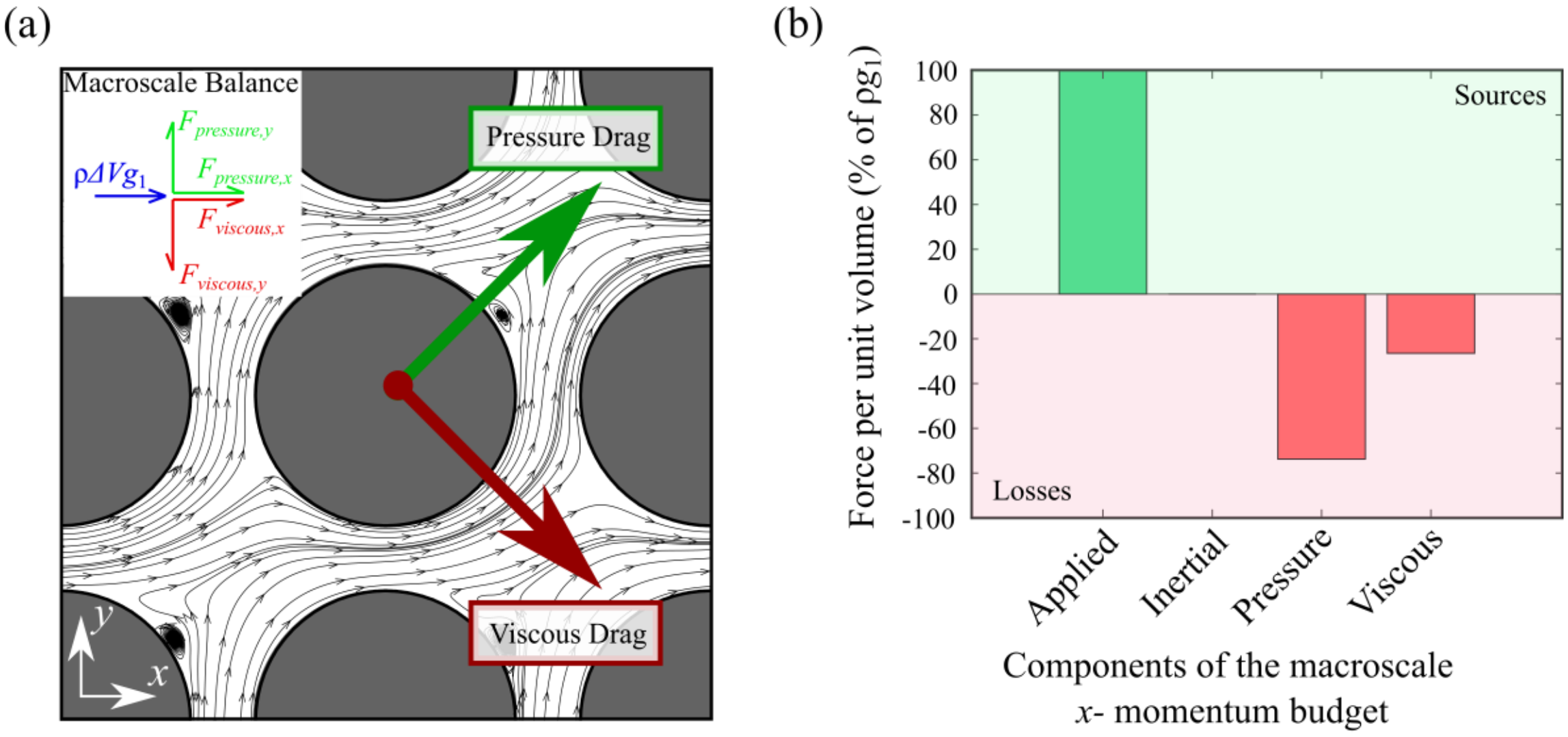

Figure 2: (a) A 2D sketch of the forces that are acting on the solid obstacle in deviatory flow (the force vectors are not to scale), (b) The macroscale $x$- momentum budget computed for the laminar case A1. The individual components of the budget are calculated in the units of force per unit volume and then normalized by the applied pressure force per unit volume $g_1$.

We will first look at the properties of macroscale turbulence before symmetry-breaking (cases A2 and A3). For the transient flow, the macroscale pressure and inertial components are dominant in the $x$- and $y$- directions when compared to the viscous component. Since the flow is symmetric, the viscous force in the $y$- direction is virtually zero. The macroscale pressure force is zero in the $z$- direction since the solid obstacles are cylindrical in shape with their axes oriented in the $z$- direction. From the macroscale perspective, the random fluctuation of the components of the budget indicate that the flow at $Re_p$ = 225 is turbulent. The order of magnitude of the forces in the $z$- direction is only one order less than that of the $y$- direction, which is indicative of three-dimensionality. We primarily study the momentum budget in the $x$- and $y$- directions since we only observe small perturbations in the $z$- direction. The development of symmetry-breaking is not studied starting from $Re_p$ = 225 because the flow after symmetry-breaking is vastly different, as shown in the following sections. Thus, the variation in the flow properties cannot be solely attributed to the symmetry-breaking phenomenon, but also to the development of turbulence.

The von Karman instability is present at $Re_p$ = 225, which introduces the dynamic behavior to the flow (Figure 3). The macroscale force is calculated by summing the forces acting on the 16 solid obstacles present in the REV-T. The pressure force acting on the individual solid obstacles and their Fourier transform (absolute values) are plotted in Figure 4. There are two features to note in the pressure force plots in the time and frequency domain – (1) the existence of two distinct, dominant time scales, and (2) the phase difference between the drag forces acting on individual solid obstacles. The phase difference can be identified in Figure 4(a) and (c), where the peaks in the pressure force on individual solid obstacles do not coincide in time. The high frequency peaks at non-dimensional frequency $f$ = 3.8 in Figure 4(b) and (d) correspond to the vortex shedding frequency. This is corroborated with the peak frequencies in $y$- velocity that are measured at the midpoint between two solid obstacles. The results are also supported by the visualization of the vortex shedding process using the streamlines and the skin-friction lines on the surface of the solid obstacle. We are using the non-dimensional frequency $f$ to analyze the time series of force components. The frequency is defined as the reciprocal of time and it is non-dimensionalized using the characteristic length and velocity. The form of the non-dimensional frequency $f$ is similar to that of the Strouhal number. The Strouhal number is used in studies of the von Karman vortex shedding frequency. In this paper, other flow instabilities are present due to the presence of an array of solid obstacles in the porous medium geometry. The porous medium geometry introduces a spectrum of frequencies of vortex shedding even at low Reynolds numbers. It is not common practice to use the Strouhal number for analysis in these conditions. However, note that the value of $f$ at the vortex shedding frequency in some cases (example: case A2) is similar to the Strouhal number. Note that the Strouhal number does not take a constant value in porous medium flows at a moderate Reynolds number, unlike the flow around a single circular cylinder. Therefore, there is no need to use the Strouhal number in this paper.

The low frequency peak is observed at two different values of frequency in the $x$- and $y$-directions ($f = 0.3$ and $f = 0.6$). The discrepancy is attributed to the finite sample size of the signal used for the Fast Fourier Transform and the peaks are, therefore, associated with the same phenomenon. The low frequency oscillations arise from a secondary flow instability. The plots of the pressure force in the frequency domain (Figure 4(b) and (d)) are qualitatively similar to the velocity probe measurements that are reported by Agnaou *et al.* (2016). The low frequency peak is a derivative of a Hopf bifurcation in periodic porous media. Since the porosity is low in this case ($\varphi = 0.5$), there exists a strong interaction between the primary and secondary flows. The macroscale pressure and inertial components of the macroscale momentum budget are in competition (Figure 3), which results in the formation of the secondary flow instability. The presence of phase difference between the pressure forces acting on the solid obstacles alleviates the high magnitude of macroscale adverse pressure gradient that is introduced by the flow instability.

At higher Reynolds numbers $Re_p \geq 300$ (cases A3 to A7), the power spectrum of the pressure force is spread over a band of frequencies. Sharp peaks like those seen in Figure 4(d) for case A2 become more diffuse at high Reynolds numbers, which is a characteristic of turbulence. The pressure forces for $Re_p = 300$ and $Re_p = 489$ (cases A3 and A4) are plotted versus $f$ in Figure 5. For case A3, there are three dominant frequencies for the pressure force, one more than for case A2. The peak at $f = 0.333$ corresponds to the secondary flow instability that was present in case A2. The peak at $f = 3.331$ corresponds to the vortex shedding process. The frequency band $0.333 < f < 3.331$ is also excited as a result of turbulence. The oscillation in the dimensional pressure force has a higher amplitude in case A3 than in case A2. The higher amplitude alludes to an increased strength of the vortex shedding process, which is verified using the magnitude of vorticity associated with the vortices. In Figure 5(a), however, the non-dimensionalization of the pressure force using the applied pressure gradient shows a decrease in the magnitude due to the non-linear increase in mean drag force. For case A4, a band of frequencies is excited with no distinct peaks in the power spectrum. The vortex shedding frequency is no longer one of the dominant frequencies for the oscillation in the pressure force. The maximum power is observed in the frequency range $1 < f < 3$ in Figure 5(b), which is one order of magnitude less than the mean vortex shedding frequency. The low frequency oscillations of the flow instability that were present in case A3 are not present in case A4. The change in the dynamics of the flow from case A3 to A4 is brought about by the symmetry-breaking process. The flow after symmetry-breaking has only one stagnation point on the solid obstacle, as opposed to the two stagnation points observed in symmetric flow. After symmetry-breaking, the stagnation point is not formed by the incidence of a vortex with the solid obstacle (see section 3.1.2).

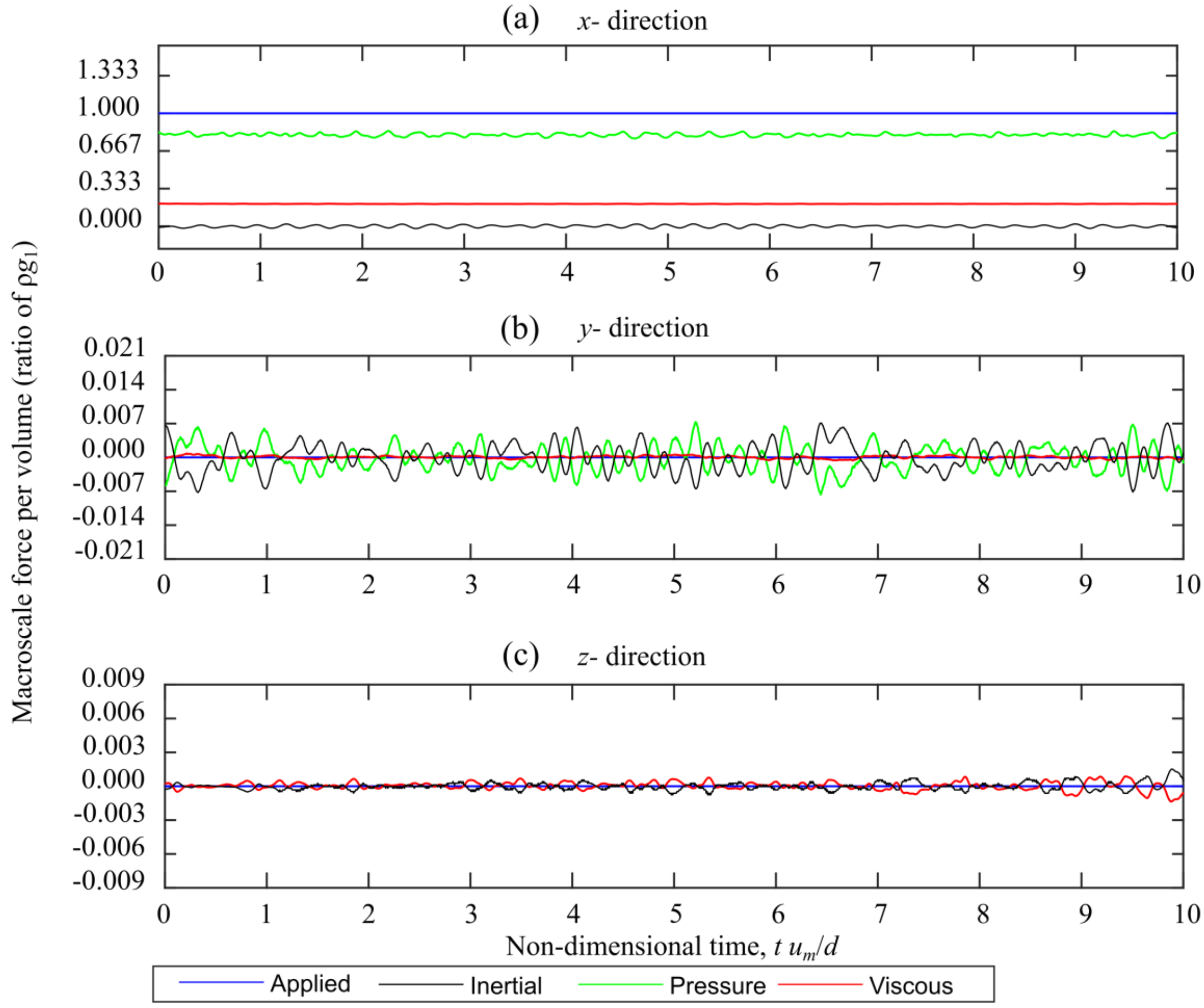

Figure 3: The dynamic macroscale momentum budget for case A2 (turbulent, $Re_p = 225$). The budget is computed for the (a) *x*- , (b) *y*- , and (c) *z*- directions. The time is non-dimensionalized using $d/u_m$, and the force components using the applied force component.

We want to make a note about some observations about the phase difference in the drag forces between individual solid obstacles. The phase difference in the drag force has an influence on the macroscale flow field in a manner similar to wave interference phenomena. It causes a reduction in the amplitude of oscillation of the total drag force inside the REV. The phase difference will also influence the definition of the macroscale Reynolds Stress terms used in turbulence modeling. Although the order in which volume and Reynolds averaging are applied to dependent quantities in the governing equations does not change the result (Pedras & de Lemos 2001), the order in which these two averaging procedures are applied to model parameters, such as the macroscale turbulence kinetic energy, may change the result. The phenomenon provides a physical reason for why the two definitions of the macroscale turbulence kinetic energy that are used in porous medium turbulence modeling are not interchangeable (reported in de Lemos 2012 and Pinson *et al.* 2006). The two commonly used definitions are $\langle\langle u'_i u'_i\rangle/2\rangle^i$ and $\langle\langle u'_i\rangle^i\langle u'_i\rangle^i\rangle/2$. The prime denotes a fluctuation term obtained from Reynolds averaging. In the first definition, the square of the velocity fluctuation is time averaged and then volume averaged. In the second definition, the velocity fluctuation is volume averaged, squared, and then time averaged. By changing the order in which these operations are applied, we are able to include the influence of the phase difference in the second definition. The macroscale turbulence kinetic energy calculated using the second definition can be substantially lower than that of the first definition depending on the magnitude of the phase difference. This argument can be extended to other terms of the Reynolds stress tensor as well. Even though the second definition of macroscale turbulence kinetic energy contains the information about the phase difference, the first definition is more commonly used for modeling. Therefore, we use this definition later in section 3.2. This feature of microscale porous media flow is a strong proponent in favor of the use of LES and DNS methods for simulation over RANS. The presence of phase difference also validates the need for large REVs to simulate turbulent porous media flows. The use of a domain with a single solid obstacle will only simulate a special case where all of the vortex systems are acting in phase. The size of the REV-T should be sufficiently large that the influence of phase difference becomes invariant.

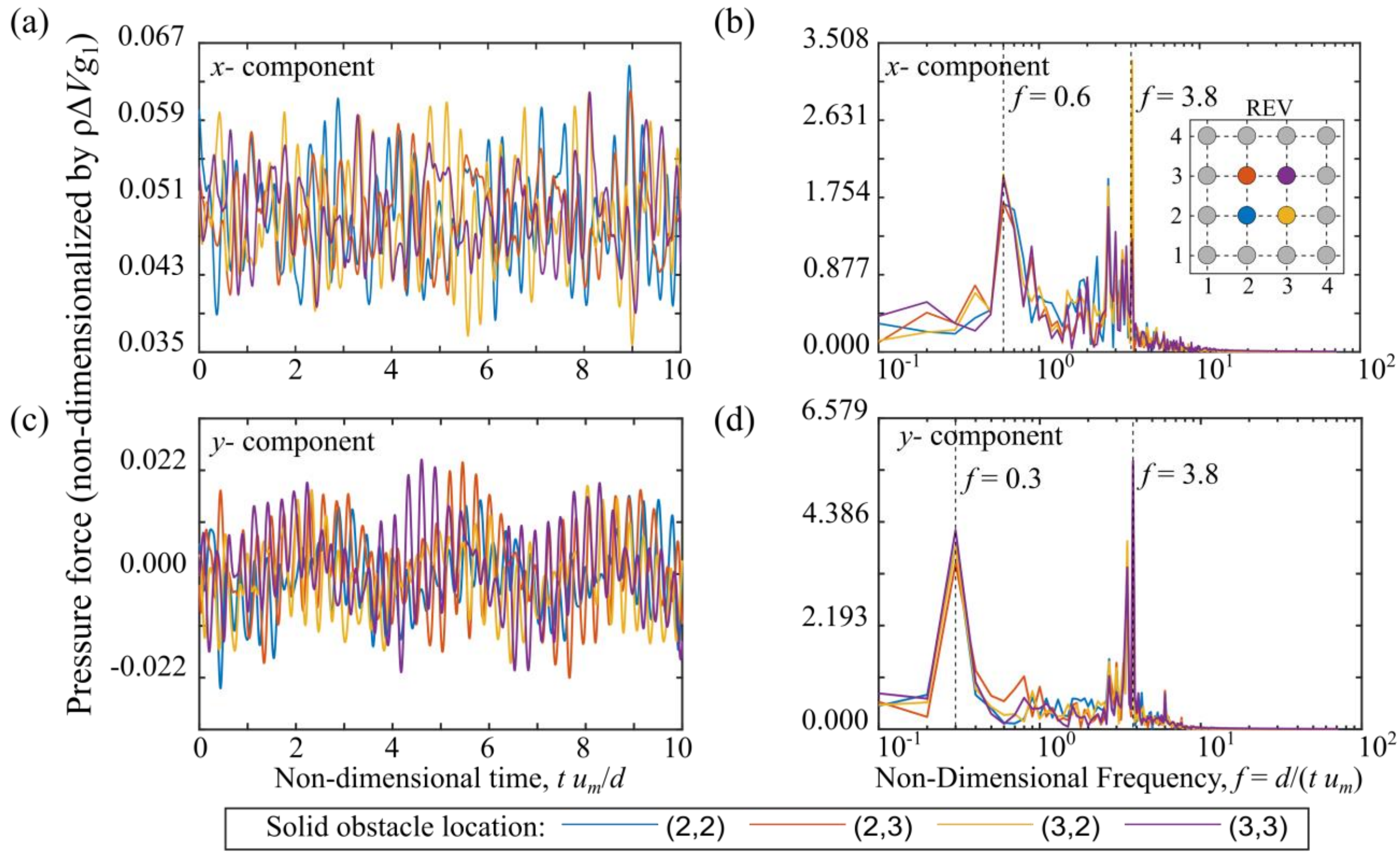

Figure 4: The pressure force acting on individual solid obstacles inside the REV-T for case A2 (Table 1) in the *x*- direction versus non-dimensional (a) time and (b) frequency, and in the *y*- direction versus non-dimensional (c) time and (d) frequency. See pressure drag definition in equation 3.1. The colors represent the location of the solid obstacle in the $4 \times 4$ matrix. Example: (2,3) refers to the solid obstacle in second place in the *x*- direction and third place in the *y*- direction.

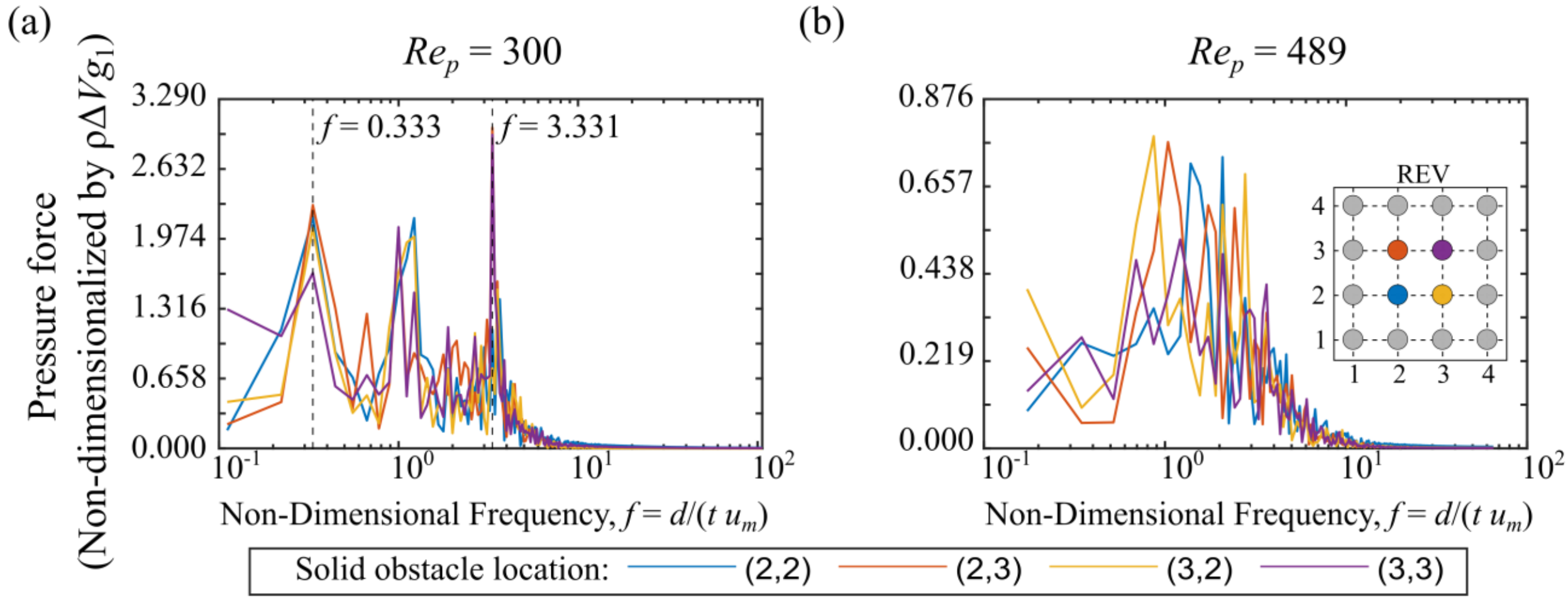

Figure 5: The pressure force acting in the *y*- direction on the individual solid obstacles inside the REV-T versus non-dimensional frequency for (a) $Re_p = 300$ (case A3), and (b) $Re_p = 489$ (case A4).

To summarize the discussion about the transition from symmetric to deviatory flow, both a change in the flow topology and the flow dynamics are expected. From the observed change in the power spectrum before and after symmetry-breaking, the underlying physics of the phenomenon is related to the following:

- The presence of a flow instability before symmetry-breaking that causes large-scale pressure oscillations and its disappearance after symmetry-breaking.
- The increase in the amplitude of the oscillation of pressure until symmetry-breaking occurs, followed by a decrease.
- The reduction in the dominance of the vortex shedding process on the pressure force after symmetry-breaking.

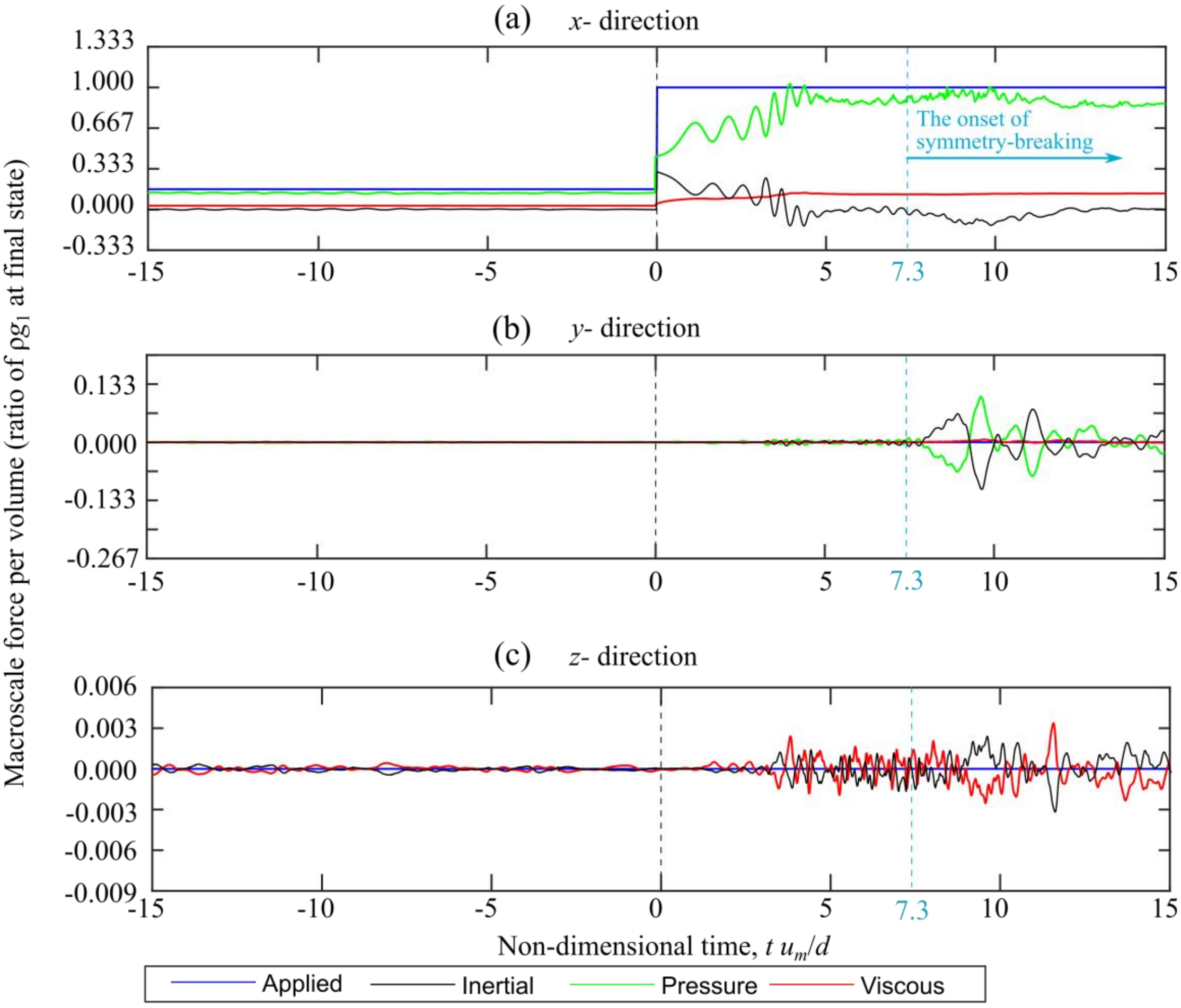

Figure 6: The macroscale momentum budget for the symmetry-breaking process ($Re_p$ = 300 to 489). The budget is computed for the (a) *x*- , (b) *y*- , and (c) *z*- directions. The time axis is shifted such that the step change in applied pressure gradient occurs at $t = 0$.

The results suggest that the symmetry-breaking is brought about by the amplification of the flow instability with the increase in Reynolds number, which ultimately leads to the breakdown of symmetry. To confirm this, the transient stages of symmetry-breaking are simulated using the following methodology. At the beginning of the simulation, the applied pressure gradient is maintained at a constant value such that the volumetric flow rate corresponds to a Reynolds number of 300. Then, a step function is used to change the applied pressure gradient at $t\, u_m/d = 0$ to that of a Reynolds number of 489. The constant value of applied pressure gradient that sustains the Reynolds number of 500 is determined a priori to use in the present simulation. The discrepancy between the prescribed Reynolds number ($Re_p = 500$) and the actual Reynolds number ($Re_p = 489$) is a result of the pathway that the bifurcation takes in this simulation. The step function is used to avoid any oscillations that will be introduced by using a control system to maintain the flow rate. This enabled the study of the transient stages towards deviatory flow without having to account for the large-scale noise that would have come from the time-advancement algorithm. However, it is not possible to segregate the momentum contribution of the symmetry-breaking phenomenon and the change in applied pressure gradient, since they occur concurrently. Therefore, the analysis of the macroscale momentum budget is supplemented by the inspection of the microscale flow field (section 3.1.2) to confirm the observations.

All of the components of the macroscale momentum budget at $t\, u_m/d = 0$ amplify in magnitude as a result of the increased supply of momentum (Figure 6). The three-dimensionality of the flow increases as a result of the increase in the Reynolds number (see Figure 6(c)). However, the magnitude of the forces in the *z*- direction are two orders of magnitude less than in the *x*- and *y*- directions. It is sufficient to take only the *x*- and *y*- directions into consideration for this macroscale analysis, while bearing in mind that the microscale flow is 3D. Both the macro- and micro- scale turbulence are strongly anisotropic, such that the streamwise direction is dominant. Before symmetry-breaking, the streamwise direction is aligned with the *x*- direction. After symmetry-breaking, the deviatory flow results in a macroscale flow angle ($\theta_{macro}$) between the streamwise direction and the *x*- direction in the *xy*- plane. The macroscale flow angle $\theta_{macro}$ is calculated as the deviation of the direction of the macroscale velocity vector from the *x*- direction.

Among the components of the macroscale momentum budget, the macroscale inertial and pressure forces dominate the dynamics of the macroscale flow. The viscous force is essential for the formation of the vortices and the flow instabilities, but it is not the driving component in symmetry-breaking. In the *x*- direction, the change in the applied pressure gradient ($\rho g_i$) leads to an increase in the macroscale *x*- pressure force and its amplitude of oscillation (Figure 6(a)). The macroscale *x*- pressure force does not grow after $t\, u_m/d = 5$ onwards as long as the applied pressure gradient is unchanged. The amplitude of oscillation of the macroscale *x*- pressure force decreases after $t\, u_m/d = 4$, which is followed by the onset of macroscale symmetry-breaking at $t\, u_m/d = 7.3$ (see macroscale *y*- direction pressure force in Figure 6(b)). In the *y*- direction, the change in the applied pressure gradient increases the amplitude of oscillation of the macroscale *y*- pressure force. Small amplitude oscillations in the macroscale *y*- pressure force are first observed in $0 < t\, u_m/d < 7.3$ as a direct result of the change in applied pressure gradient. Large amplitude oscillations in the macroscale *y*- pressure force are observed in $t\, u_m/d > 7.3$ because of the deviatory flow from symmetry-breaking. The change in the amplitude of oscillation of the pressure force due to symmetry-breaking is more prevalent in the *y*- direction (Figure 6(b)) than in the *x*- direction (Figure 6(a)). In the *y*- direction, amplitude of oscillation of the pressure force increases by one order of magnitude due to the deviatory flow. Therefore, the transformation of the microscale flow field in the *xy*- plane after symmetry-breaking is more evident in the *y*- direction macroscale momentum budget than in the *x*- direction.

The macroscale *y*- pressure force oscillates about zero both before and after symmetry-breaking. This is counterintuitive to the observations in Figure 2(a), where a non-zero *y*- pressure force is balanced by a non-zero *y*- viscous force in order to sustain the deviatory flow. If the components of the *y*- direction macroscale momentum budget have a zero mean value after symmetry-breaking, how can the deviatory flow solution exist? To answer this question, the pressure forces that are acting on the individual solid obstacles are examined (Figure 7). The $4 \times 4$ matrix of solid obstacles is indexed by dividing the matrix into rows and columns. The rows are aligned with the *x*- axis and the columns are aligned with the *y*- axis.

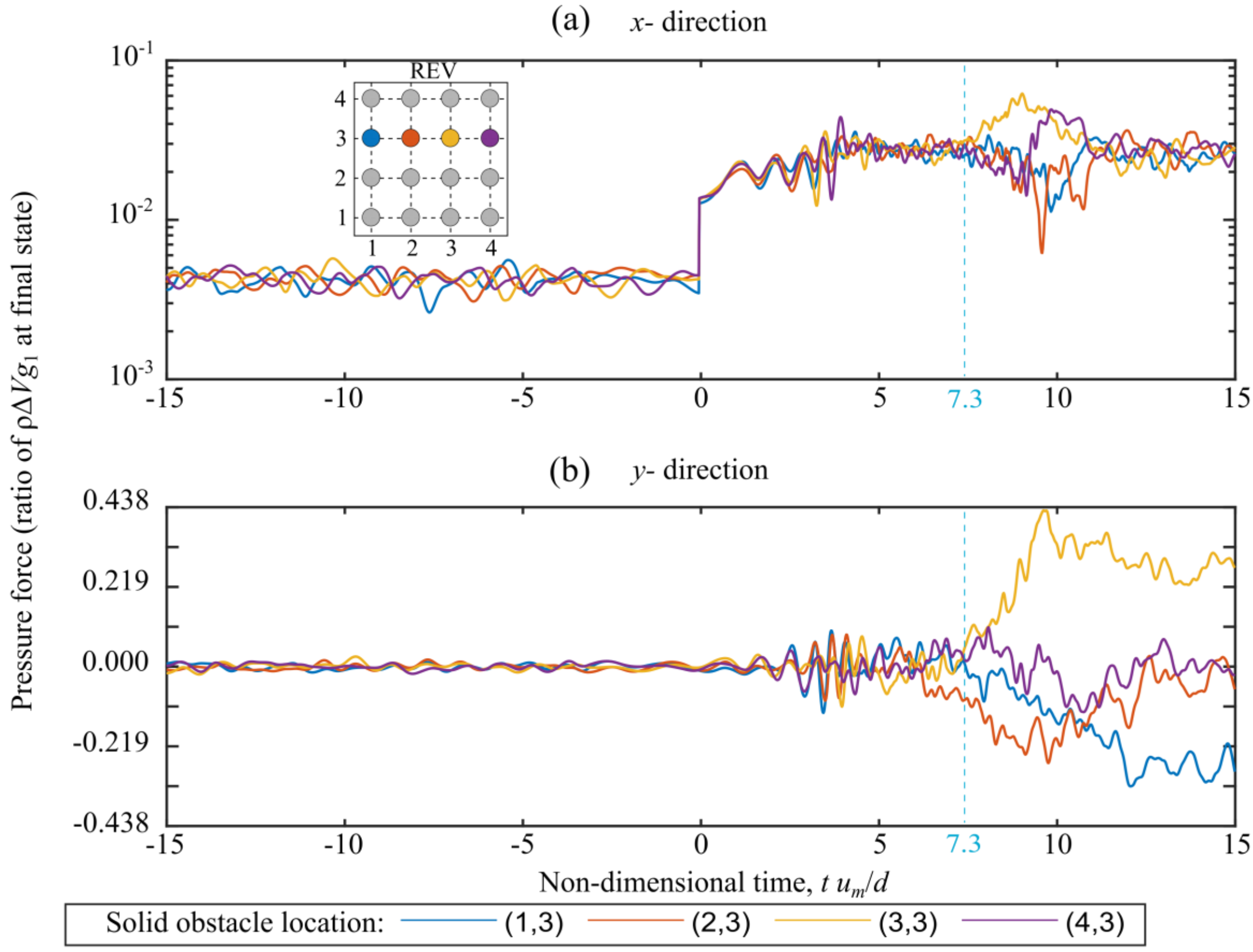

Figure 7: The pressure forces acting in the (a) *x*- and (b) *y*- directions on the individual solid obstacles inside the REV-T versus time for the transient simulation ($Re_p$ = 300 to 489).

After symmetry-breaking has developed ($t\, u_m/d > 10$), the *y*- pressure forces on the individual solid obstacles in a single row have different magnitudes and sign. Two solid obstacles in the row have a positive *y*- pressure force and the other two have a negative *y*- pressure force, which results in a zero sum. The *y*- pressure force on all of the solid obstacles in a single column have similar magnitudes and sign. We noted earlier that the *y*- pressure force on the solid obstacle can be positive or negative for a corresponding positive or negative macroscale flow angle. Since there are 4 columns of solid obstacles in this REV-T, the deviatory flow behind each solid obstacle can assume either solution behind each column. In this paper, each unique combination of the *y*- pressure forces on the individual solid obstacles is called a mode, analogous to wave modes. Modes are formed in the *x*- direction alone, along which the pressure gradient is applied. The modes of deviatory flow need not be symmetric in the *x*- direction. The modes are illustrated in section (3.1.2). If the REV-T consists of one solid obstacle, only a single mode of the symmetry-breaking phenomenon with a unidirectional *y*- pressure force can be formed. However, the case of unidirectional deviatory flow is only a subset of the possible solutions in periodic porous media. The number of possible modes of deviatory flow is decided by the number of solid obstacles in the REV in the direction of applied pressure gradient. This confirms the requirement that the size of the REV-T must be greater than one pore size.

The onset of macroscale symmetry-breaking is evident in the *x*- pressure forces at $t\, u_m/d = 7.3$ (Figure 7(a)). Since the magnitude change of the *x*- pressure force at $t\, u_m/d = 0$ is large, the amplitude of oscillation of the *x*- pressure force is small in relation. Therefore, the *y*- pressure forces are used to analyze the symmetry-breaking phenomenon behind individual solid obstacles. There exists a phase difference between the pressure forces on the individual solid obstacles (Figure 7). The onset of microscale symmetry-breaking is marked by the first increase in the $y$- pressure force on the solid obstacles. The flow around the solid obstacle at location (2,3) is the first to deviate from symmetric behavior (Figure 7(b)). The deviation occurs sooner in the microscale ($t\, u_m/d = 6$) than it does in the macroscale ($t\, u_m/d = 7.3$), implying that symmetry-breaking begins as a localized, microscale phenomenon.

The flows around the individual solid obstacles break symmetry at different times because of the phase difference in the vortex shedding. The solid obstacle at location (2,3) is defined to have the leading phase at the time of microscale symmetry-breaking. The increasing order of phase lag for the solid obstacles in the third row of the REV-T is the following: (2,3) < (3,3) < (1,3) < (4,3). This order is identified by determining the solid obstacle locations where the $y$-pressure force will peak next after the solid obstacle at (2,3). The order in which the flow around each solid obstacle breaks symmetry is the same as that of the phase lag. Thus, the phase lag determines the time at which symmetry-breaking occurs for each solid obstacle. Therefore, the phase of the vortex shedding cycle at which the flow breaks symmetry is different for each solid obstacle. The phase difference is only partly responsible for the formation of the different modes. It is shown in section (3.1.2) that the flow field around the neighboring solid obstacle plays a role in the formation of the modes as well.

The formation of the phase difference and the randomness in the vortex motions behind the solid obstacles suggests the independence of the microscale flow behind each solid obstacle. The formation of modes of deviatory flow along the rows of solid obstacles suggests the dependence of the microscale flow behind a solid obstacle on the flow around the neighboring obstacles. The flow patterns around the solid obstacles in the direction of the applied pressure gradient are not identical, but the flow patterns are identical in the normal direction.

#### *3.1.2. Microscale flow field and turbulent structure visualization*

The macroscale momentum budgets for cases A1-A4 in Table 1 revealed that the flows before and after symmetry-breaking are dominated by the pressure forces. The dynamic behavior of the flow is determined by the micro-vortices, formed as a result of the viscosity of the fluid. However, the viscous force has little influence in the macroscale momentum budget. In this section, the microscale flow field is visualized to connect the observations of the macroscale flow to the microscale flow physics. The laminar flow patterns at $Re_p = 100$ are similar to those in Figure 1(a), with distinct primary and secondary flow regions. The secondary flow region consists of a recirculating (attached) vortex system. Upon transition to turbulence, three-dimensional features appear in the microscale flow at $Re_p = 225$, consistent with the macroscale momentum budget in Figure 3. Even though the micro-vortices at $Re_p = 225$ and at $Re_p = 100$ appear similar, they begin to deform in the $z$- direction at $Re_p = 225$ due to the vortex stretching process in turbulence. The 2D turbulent structures in Figure 8(a) are the micro-vortices, which possess a swirling characteristic. Note that we are using the term $\rho g_1 \Delta V$ to scale the microscale pressure distribution to be consistent with the definition of pressure drag force in equation 3.1. The pressure drag force can be calculated from the pressure distribution by integrating the microscale pressure distribution over the area of the solid obstacle.

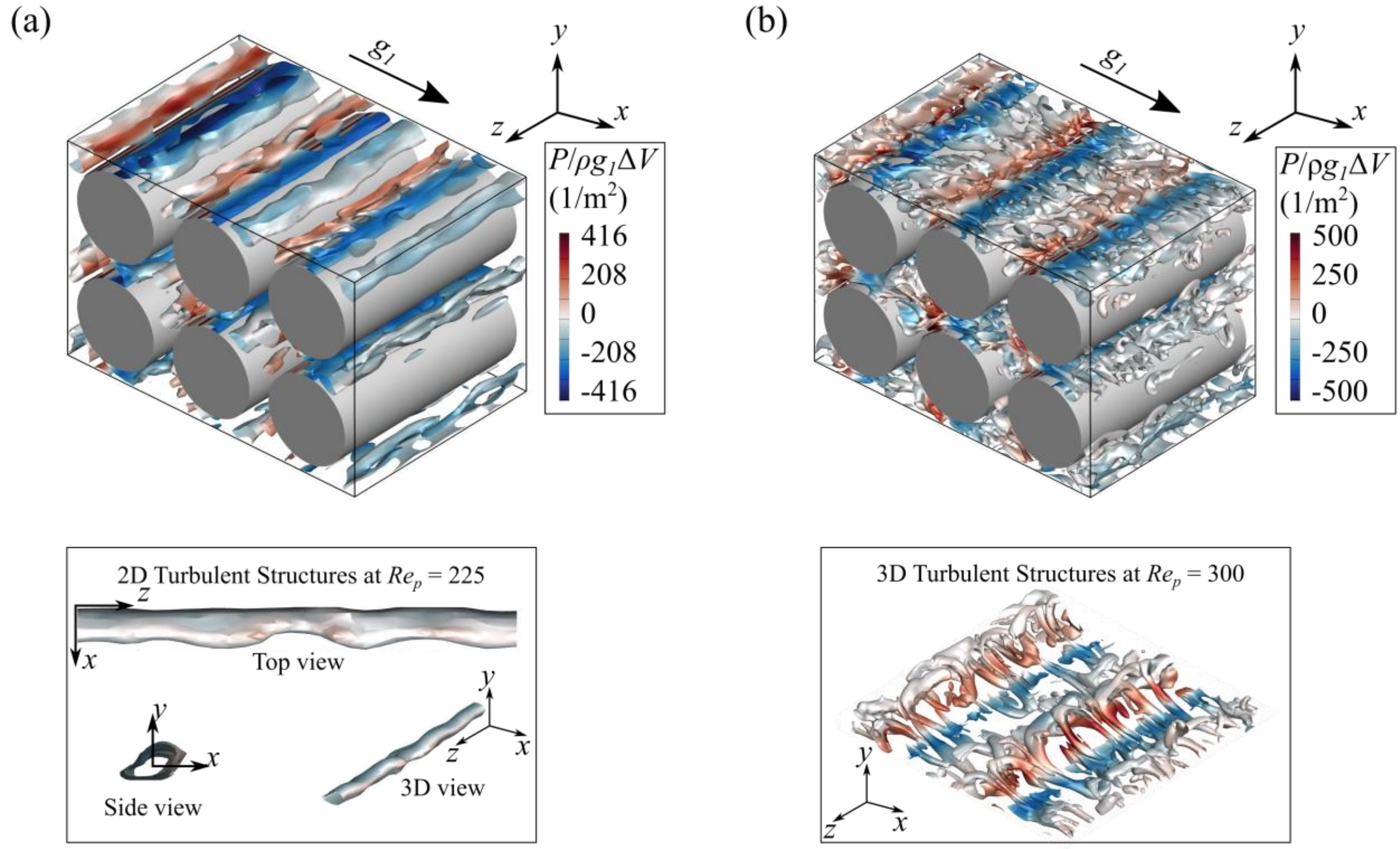

Figure 8: Coherent turbulent structures visualized using the iso-surfaces of the Q- criterion for (a) $Re_p = 225$ (case A2), and (b) $Re_p = 300$ (case A3). The Q- criterion is normalized using the maximum value and the iso-surfaces of $Q/Q_{max} = 0.001$ are plotted. A sub volume of dimensions (3*s*, 2*s*, 2*s*) of the REV-T is shown in these figures.

The micro-vortices at $Re_p = 225$ evolve in time due to turbulence production and dissipation. The micro-vortices are stationary in the void space between two solid obstacles. The micro-vortices interact with the primary flow to form flow stagnation points on the solid obstacle surface at the converging portion of the GPM. This leads to the formation of the secondary flow instability. The converging geometry intrinsically introduces a local streamwise favorable pressure gradient. However, the flow stagnation reduces the favorability of the pressure gradient in the converging section, which is the source of macroscale pressure drag. The ratio of the stagnation pressure to the applied pressure gradient increases with the Reynolds number. The presence of a substantial stagnation pressure in the converging region supports the notion that the flow instability should arise from the competition between the pressure and inertial components of the macroscale momentum budget.

There is a stark contrast between the turbulent structures observed at $Re_p = 225$ (Figure 8(a)) and $Re_p = 300$ (Figure 8(b)). At $Re_p = 225$, semi-infinite 2D turbulent structures with 3D deformations are visible. The turbulent structure has infinite dimension in the *z*- direction because of the periodic boundary condition. The deformations that are present in the turbulent structures arise from turbulent vortex stretching. At $Re_p = 300$, the turbulent structures are finite, three-dimensional and they assume the shape of hairpin vortices. The hairpin vortices are oriented in a reverse direction when compared to the observations in classic external flows like turbulent boundary layer flow (Eitel-Amor *et al.* 2015). Unlike the boundary layer flow, the head-to-tail direction of the hairpin vortex aligns with the streamwise direction of the flow. Reverse hairpin vortices have also been observed in periodic porous media by Jin & Kuznetsov (2017) for spherical solid obstacles. To understand the unique vortex shape, the microscale turbulent structures are classified into the following based on the swirl – the micro-vortices and the turbulent eddies. The regions of swirl-dominated flow are identified by determining the center of swirling flow using the algorithm developed by Sujudi & Haimes (1995). The micro-vortices are characterized by swirling flow. The turbulent eddies do not possess a swirling vortex core. The turbulent structures that are observed in the Q- criterion plots (Figure 8) in the regions without vortex core lines (Figure 9(a)) are turbulent eddies.

The direction of the hairpin vortex structure is dependent on the microscale spatial inhomogeneity in turbulence dissipation, brought about by distinct primary and secondary flow regions. At $Re_p = 300$ (case A3), the micro-vortex core lines are concentrated in the secondary flow region (Figure 9(a)). The swirling motion of the micro-vortex is localized in the secondary flow region. The primary and secondary flow regions are separated by a region of strong shearing flow (see vorticity contours in Figure 9(a)). In order to understand the reversed direction of the hairpin vortices, consider an unperturbed vortex filament that is present in the secondary flow region (Figure 9(b)). The micro-vortex is perturbed and stretched in the $z$-direction by turbulence. As it stretches, the perturbed micro-vortex is subjected to the swirling secondary flow region and the highly dissipative primary flow region simultaneously. The low velocity and the rotational nature of the secondary flow region is favorable for the sustenance of the micro-vortex. The high rate of strain and the presence of a pressure gradient in the primary flow region are adverse to the sustenance of the micro-vortex. Therefore, the micro-vortex breaks up in the primary flow region, while it continues to sustain in the secondary flow region. This results in the formation of the reverse hairpin vortex with the head in the secondary flow region and the tail in the primary flow region.

The head of the reverse hairpin vortex will reduce in size as the vortex is gradually weakened due to energy dissipation in the flow. The phase difference in the micro-vortex motions behind the different solid obstacles is visualized by comparing the size of the head of the reverse hairpin vortices. In Figure 8(b), the 3D turbulent structures for the two adjacent solid obstacles show that the head of the reverse hairpin vortex is larger for the upstream vortices. Therefore, the upstream vortices are lagging behind resulting in a phase difference in the vortex motions. This observation is supported by the fact that the upstream vortices have a lower magnitude of pressure associated with them when compared to its neighbor. The phase difference in the pressure forces was also observed in Figure 7(b). The reverse hairpin vortices in Figure 8(b) possess some degree of order in their arrangement, forming a “knit” pattern of 3D turbulent structures. The formation of the pattern of similar turbulent structures that arise from the micro-vortices is consistent with the formation of significant peaks in the frequency distribution of the pressure forces (Figure 5(a)). The micro-vortex patterns behind all of the solid obstacles have a similar size, shape, and turn-over time. The summation of the influence of an infinite number of these vortices on the solid obstacles leads to the excitation of the pressure force at the frequency of vortex shedding. The inherent randomness in the micro-vortex pattern is responsible for the excitation of the other frequencies.

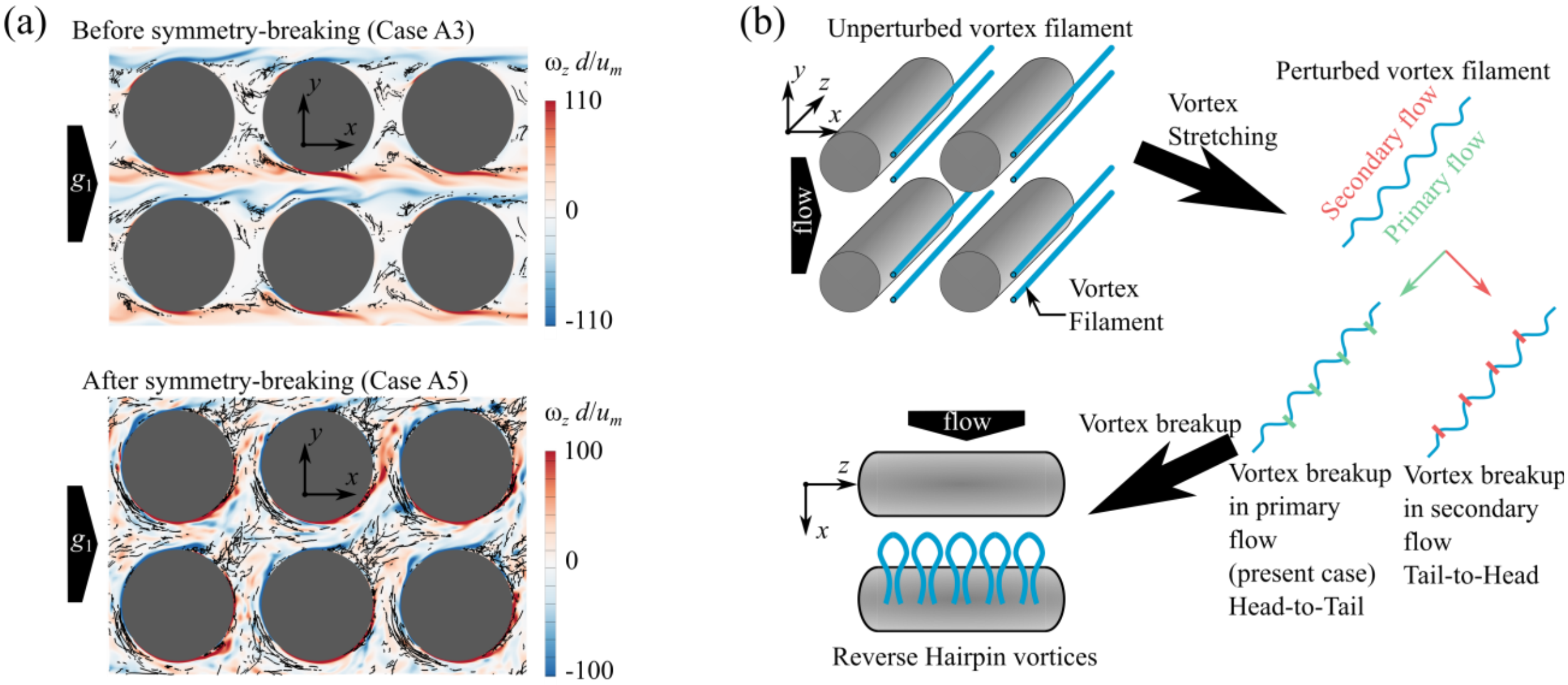

Figure 9: (a) Instantaneous vortex core lines before and after symmetry-breaking projected on a plane at $z$ = 0 overlaid on contours of $z$- vorticity. (b) A sketch illustrating the process of formation of the reverse hairpin vortices in porous media.

Before symmetry-breaking, micro-vortices are not transported beyond a single pore space (Figure 9(a)). The micro-vortices transform into turbulent eddies in the shear flow region. These turbulent eddies are observed to move downstream beyond a single pore space while continuing to diminish in size. The turbulent eddies have a random shape that does not bear resemblance to the reverse hairpin vortices. On this basis, the domain of influence of the micro-vortex motion behind one solid obstacle is limited to its neighbors. An indirect influence is present from the propagation of turbulent eddies until they are completely dissipated. After symmetry-breaking, the distinction between the primary and the secondary flow regions is lost, as is the distinction between regions of micro-vortices and turbulent eddies. The flow separation point advances to a downstream location resulting in the formation of a small vortex wake and a weak shear layer. The micro-vortices propagate downstream and transform into turbulent eddies under the influence of the near-wall dissipation and the pressure gradient from the solid obstacle geometry. The degree of confinement of the turbulent structures in the pore space is reduced by the occurrence of symmetry-breaking evidenced by the micro-vortices travelling a greater distance before completely dissipating.

At $Re_p$ = 489, the deviatory flow results in the formation of fine-scale turbulent structures (Figure 10), when compared to $Re_p$ = 300. Fine-scale turbulent structures are formed because of the reduction in the width of the micro-vortex wake after symmetry-breaking. Fine-scale turbulent structures are also formed because of increased shredding of the turbulent structures at higher Reynolds numbers (Wood *et al.* 2020). The “knit” pattern in the micro-vortices (Figure 8(b)) is no longer present due to increased turbulence intensity. This is reflected in the loss of dominance of the vortex shedding frequency on the pressure force and the distribution of the pressure force across a band of frequencies (Figure 5(b)). The overall distribution of turbulent structures is chaotic. The deviatory flow is not evident from the Q- structures because of the formation of a mode of deviatory flow that increases the tortuosity of the streamlines (Figure 10(b)). For this reason, the visualization of 3D Q- structures is supplemented by the visualization of 2D projections of the flow streamlines. The streamline plot shows that the flow around each solid obstacle consists of a single stagnation point and a detached vortex system. Separation bubbles are formed when the location of the stagnation point changes between adjacent solid obstacles. The low-pressure turbulent structures formed in the detached vortex system (at obstacles 1 and 4 in Figure 10(c)) are micro-vortices with a vortex core line inside of them. The elongated, high-pressure turbulent structures near the stagnation point (at obstacles 2 and 3 in Figure 10(c)) are micro-vortices that are impinging from the upstream neighboring solid obstacle, marking the beginning of their transformation to turbulent eddies.

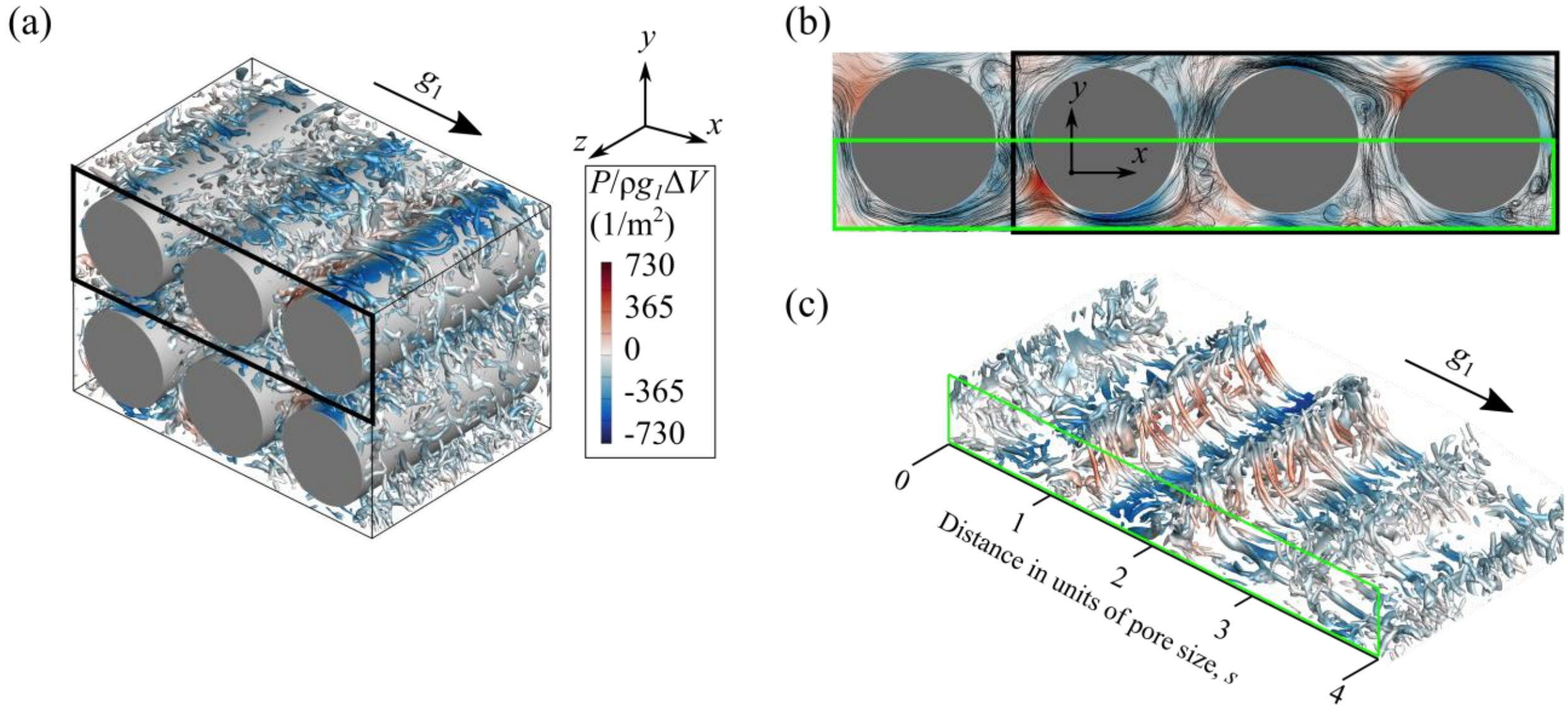

Figure 10: (a) Coherent turbulent structures visualized using the iso-surfaces of the Q- criterion for $Re_p = 489$ (case A4). The Q- criterion is normalized using the maximum value and the iso-surfaces of $Q/Q_{max} = 0.02$ are plotted. A sub volume of dimensions (3*s*, 2*s*, 2*s*) of the REV-T is shown in these figures. (b) Instantaneous 3D flow streamlines projected on a plane at $z = 0$ overlaid on contours of static pressure. (c) Coherent turbulent structures visualized using the Q- criterion for a sub volume of dimensions (2*s*, 0.5*s*, 2*s*).

The 3D turbulent structures observed in Figure 10(a) and (c) after symmetry-breaking bear no resemblance to the reverse hairpin structures and the "knit" pattern observed before symmetry-breaking (Figure 8(b)). The reverse hairpin structures are formed by the vortex stretching of a 2D vortex structure that is semi-infinite in the *z*- direction at the time of formation. A flow separation line is formed behind the solid obstacle that is virtually one-dimensional (Figure 11(a)). After symmetry-breaking ($Re_p = 489$), the size of the vortex core decreases and fine-scale structures are formed. The flow separation line after symmetry-breaking is more complex (Figure 11(b)). The flow separation lines are 2D and discontinuous with local flow re-reattachment, which together result in the formation of finite 3D vortex structures. The micro-vortices are the primary source of turbulent structures in periodic porous media. Therefore, the difference in the large-scale turbulent structures before and after symmetry-breaking results in the change in the turbulence intensity and degree of anisotropy.

The origin of microscale symmetry-breaking can be traced from the following differences in the microscale flow before and after symmetry-breaking:

1. The non-linear increase in the microscale flow stagnation pressure with increase in the Reynolds number.

2. The change in the location and the number of flow stagnation points. The change in the location of the flow separation points and the micro-vortex core size. The formation of fine-scale turbulent structures as a result of the change in micro-vortex size.
3. The disappearance of the order in the flow patterns after symmetry-breaking and the increased turbulence intensity and dissipation in the microscale flow.

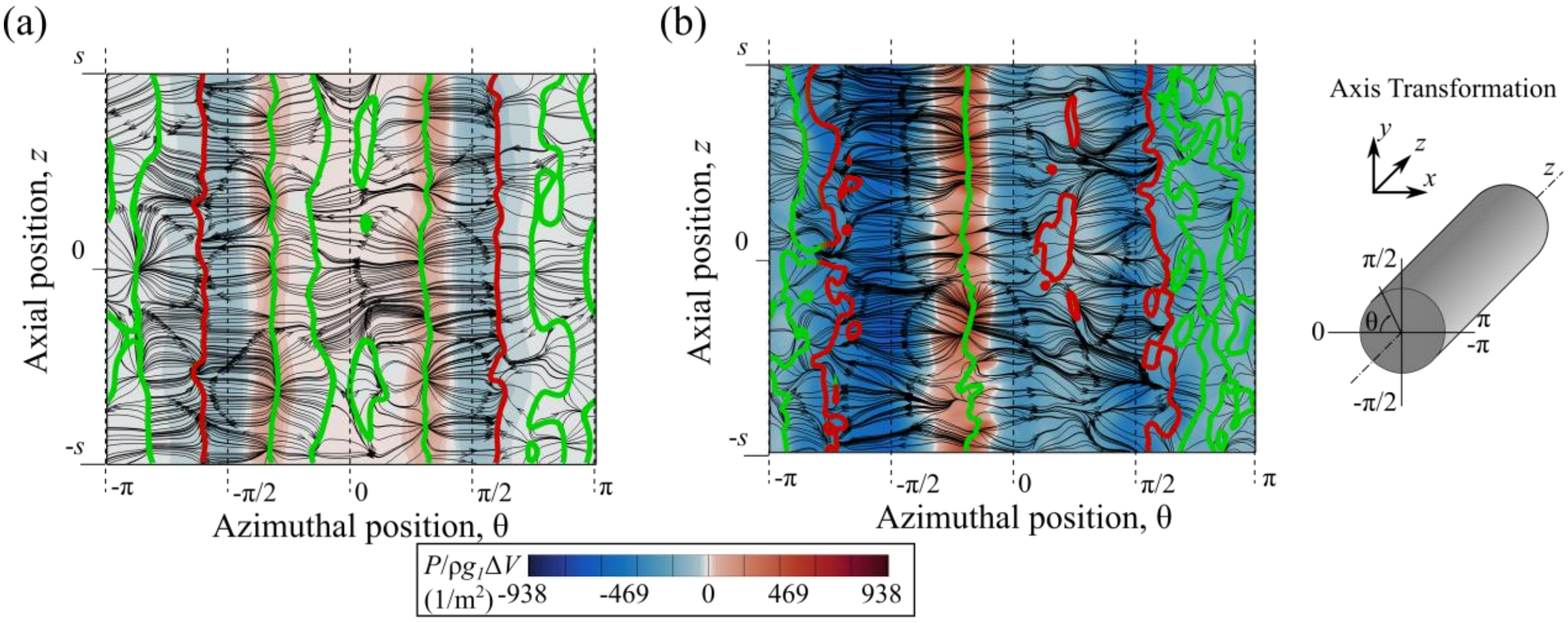

Figure 11: Skin-friction lines (black) plotted on the surface of the solid obstacles for the flow (a) before symmetry-breaking, and (b) after symmetry-breaking. The red and green lines are iso-lines of zero shear stress indicating the flow separation line and the vortex regions, respectively.

The transient simulation that was shown in Figure 6 is revisited to visualize the stages of symmetry-breaking. The flow field is three-dimensional for the entire duration of the simulation. The microscale flow field is visualized using a 2D projection of the instantaneous flow streamlines (Figure 12). Microscale flow structures are visualized using 3D coherent structures with the Q- criterion and 2D contour plots of vorticity magnitude (Figure 13, movies 1 & 2 available in supplementary material). The stages of symmetry breakdown are summarized in Table 2.

| Stage | Time $t\, u_m/d$ | Flow Properties |
|---|---|---|
| A | 0.00 | The applied pressure gradient is increased to change $Re_p$ from 300 to 489. A recirculating vortex is present in the secondary flow region. |
| B | 3.14 | The magnitude of microscale pressure increases in response to the applied pressure gradient. |
| C | 4.71 | The three-dimensionality of the flow increases, fine-scale turbulent structures appear. The recirculating vortex system breaks down into a recirculating system of small turbulent structures. |
| D | 7.85 | The high magnitude of stagnation pressure “plugs” the flow, resulting in lateral favorable pressure gradient locally. Deviatory flow follows the breakup of the recirculating vortex behind column 2 of the REV. |

| E | 11.00 | The flow around the remaining columns of the REV breaks symmetry, propagating away from column 2. |
|---|---|---|

Table 2: The stages of symmetry-breaking inferred from the transient simulation illustrated in figures 12 and 13.

The macroscale $x$- velocity increases when the applied pressure gradient is increased. The use of a constant value of the Q- criterion normalized by the maximum value for all of the time steps will reduce the visualization quality. Therefore, the Q- criterion is normalized using the solid obstacle diameter $d$ and the instantaneous macroscale $x$- velocity $\langle u_1 \rangle^i$. The microscale flow field at stage A is taken as the reference point before symmetry-breaking (Figure 12(a)). The flow streamlines show a large recirculation zone in the secondary flow region with a size similar to the radius of the circular cylinder solid obstacles. The 3D hairpin structures that were introduced in Figure 8(b) are visible as hairpin-shaped streaks in the vorticity magnitude plot (Figure 13 (c)). The tail length of the reverse hairpin vortices increases from stage A to B, signifying an increase in vortex deformation due to higher strain rate in stage B compared to A. The 3D nature of the flow is not apparent in the flow streamline plots, which is the case in subsequent time steps as well. However, the flow streamlines in the $xy$- plane clearly indicate the onset and development of symmetry-breaking.

Flow recirculation in the secondary flow region before symmetry-breaking forms a strong pressure “plug” in the flow. The word “plug” is in quotation marks because it is not a solid plug and therefore allows mass to flow through it. Before symmetry-breaking, two stagnation points are formed where the micro-vortex meets the solid obstacle on its top and bottom surfaces ($\theta \sim \pm\pi/3$ in Figure 11(a)). The stagnation points are located in the converging portion of the GPM ($-\pi/2<\theta<0$ and $0<\theta<\pi/2$). The stagnation point creates a region of locally adverse pressure gradient, which is followed by the intrinsically adverse diverging portion of the GPM ($\pi/2<\theta<\pi$ and $-\pi<\theta<-\pi/2$). The magnitude of stagnation pressure increases by a factor of 100 during symmetry-breaking, when the applied pressure gradient is only increased by a factor of 6 (see Figure 11(b) versus Figure 11(a)). A pressure “plug” in the flow is formed by the substantial increase in the stagnation pressure that increases the adversity of the pressure loss in the GPM. An adverse pressure gradient in the turbulent flow increases turbulence dissipation and reduces the size of the vortex structures (Lee & Sung 2008; Tanarro *et al.* 2020).

Fine-scale turbulence structures are observed at $t\, u_m/d = 4.71$ as a result of the increased adverse pressure gradient and the increase in Reynolds number (compare Figure 13 (a) and (c) to (b) and (d)). The flow streamlines do not change significantly from stage A to C, implying that the velocity distribution of the large-scale microscale flow is unchanged. Recall that the flow transition from $t\, u_m/d = 3.14$ to 4.71 is characterized by an increase followed by a decrease in the amplitude of oscillation of the $y$- direction pressure force (Figure 7). In stage C, the three-dimensionality of the flow field increases and the reverse hairpin vortex structures are no longer visible. The large-scale micro-vortex breaks down into smaller vortices that continue to possess the large-scale swirling motion. The fine-scale micro-vortices rotate about the same vortex core, seen at stage A. However, the large-scale swirling motion is weakened by the vortex breakdown and subjected to the continuous interaction of the constituent small turbulent structures.

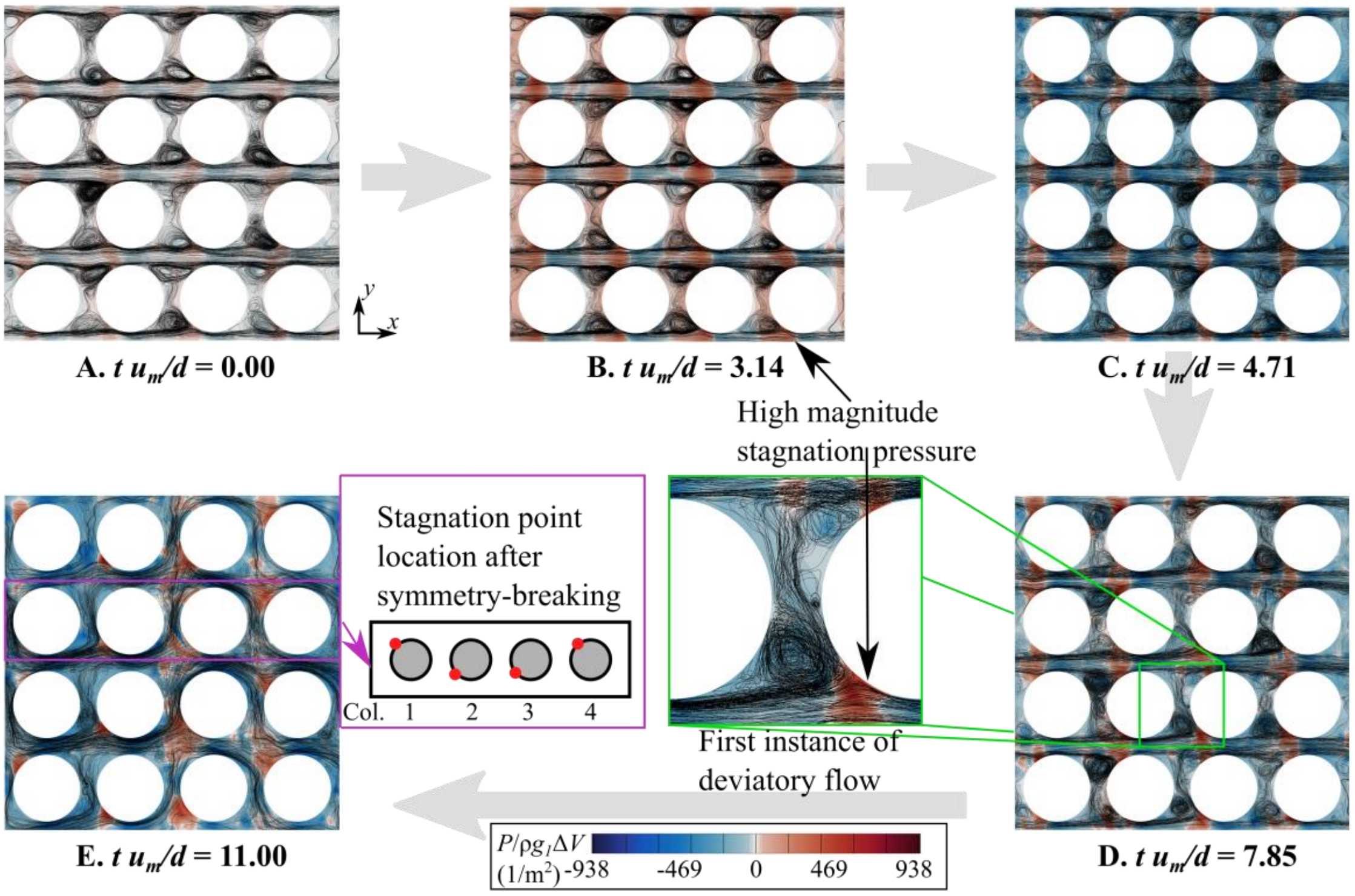

Figure 12: The transient stages A – E of the symmetry-breaking process from symmetric to deviatory flow. Instantaneous flow streamlines are seeded on the plane $z = 0$ and projected on the contours of pressure on the plane at $z = -s$.

When the pressure “plug” is formed in the streamwise direction ($x$- direction), the pressure gradient in the lateral direction ($y$- direction) becomes favorable. The flow instabilities (von Karman and secondary flow instabilities) cause asymmetry in the instantaneous pressure force that is acting on the solid obstacles. The asymmetry in the $y$- direction pressure, compounded by the increase in the stagnation pressure, will introduce a sufficient lateral favorable pressure gradient to sustain lateral flow. The $y$- pressure force sustains the deviatory flow after symmetry-breaking (Figure 2(a)). Therefore, only a trigger is required to induce symmetry-breaking, after which the deviatory flow is sustained by the lateral favorable pressure gradient. At stage D (Figure 12(d)), the flow behind the solid obstacles at column 2 of the REV-T begins to deviate from the symmetric flow configuration. Symmetry-breaking is triggered by the breakup of the swirling vortex motion at the top. The vortex breakup is caused by the strong turbulent shear and the interaction between the weaker turbulent structures in the secondary flow region. The vortex breakup results in an efflux of fluid from the secondary flow region to the primary flow region. The efflux is compensated by the influx of fluid from the bottom recirculating vortex motion aided by the lateral favorable pressure gradient. The induced fluid flow in the $y$- direction triggers the symmetry-breaking process.

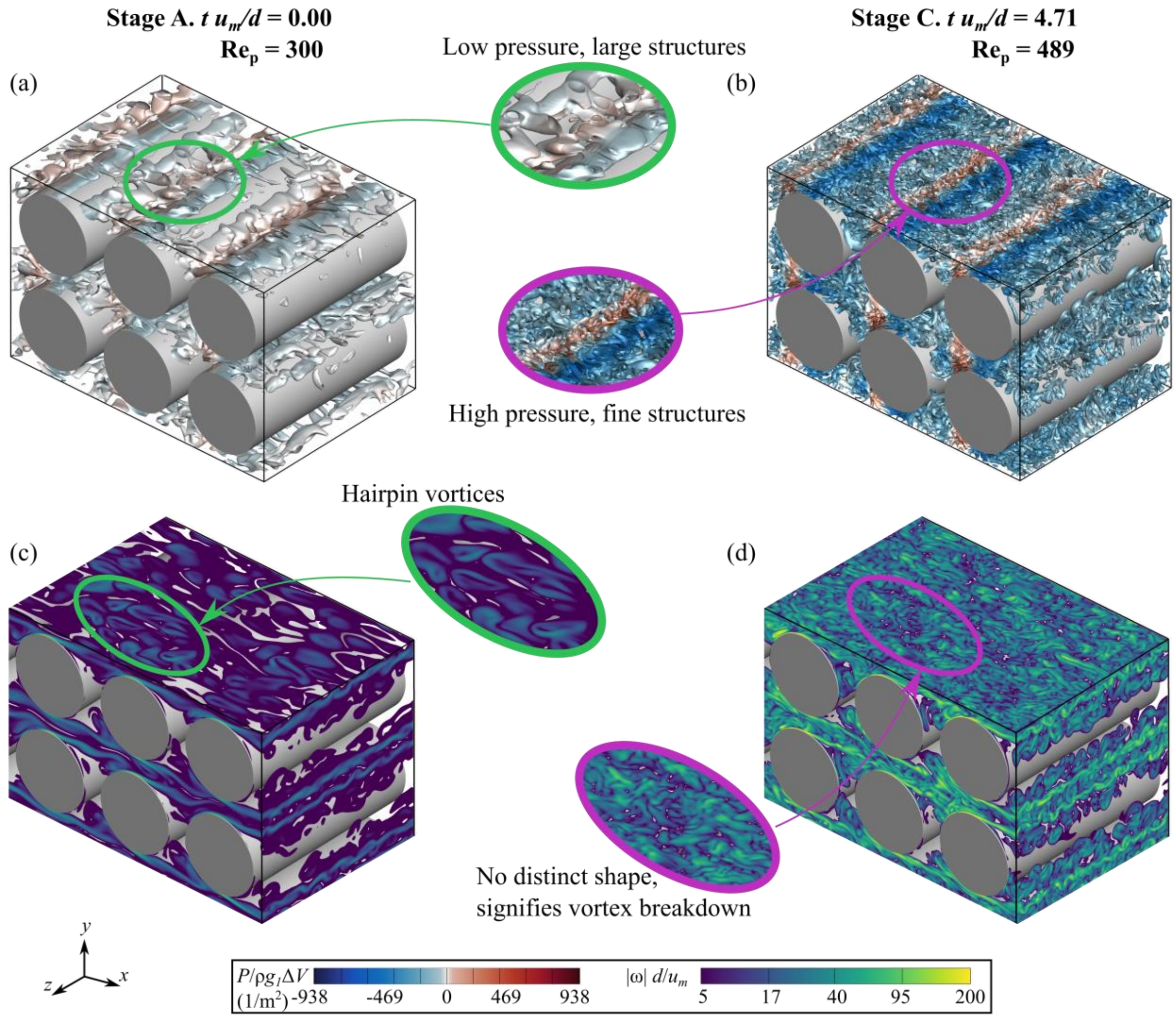

Figure 13: Transition in the 3D flow features before symmetry-breakdown due to increase in $g_i$. 3D turbulent structures colored by pressure are visualized in the sub volume (3*s*, 2*s*, 2*s*) of the REV-T using iso-surfaces of $Q(d/\langle u_1\rangle^i)^2$=50. Turbulent structures are identified by the vorticity magnitude distribution shown on the boundary of the REV-T. Vorticity contours below a magnitude of 5 are cut-off.

The information about the symmetry-breaking of the flow around solid obstacles in column 2 of the REV-T propagates to the neighboring solid obstacles. The flow around the solid obstacles in column 3, downstream of column 2, is the next to break symmetry. The flow around the solid obstacles in column 1, upstream of column 2, also changes to accommodate the deviatory flow, followed by column 4. Symmetry-breaking to deviatory flow propagates away from its location of incidence, similar to the propagation of perturbations in waves. The final mode of deviatory flow field after symmetry-breaking is reflected in the flow streamlines at stage E (Figure 12(e)). The mode is decided by the phase of the vortex motion at the time that the symmetry-breaking perturbation reaches the column of solid obstacles. The direction of the lateral favorable pressure gradient at the time chooses the direction of symmetry-breaking. This behavior is more discernable from the *y*- pressure force plots in Figure 7(b). The mode of the deviatory flow after symmetry-breaking is sustained forever, as long as the Reynolds number ($Re_p$) is maintained above the critical value for symmetry-breaking. An investigation of the modes of deviatory flow due to symmetry-breaking in porous media is beyond the scope of this paper.

### *3.2. The influence of geometry and flow parameters on symmetry-breaking*

The different modes of deviatory flow will impart a unique contribution to the macroscale flow properties. In order to understand how the microscale flow symmetry-breaking influences macroscale turbulence transport, the case of unidirectional deviatory flow is considered to reduce the complexity of the analysis. This is a mode of deviatory flow where the direction of symmetry-breaking is identical behind all of the solid obstacles. To enforce this particular mode, the deviatory flow field for an REV with only one solid obstacle is periodically repeated and interpolated onto the REV-T with the 4x4 solid obstacles. The flow solution is equilibrated and then sampled for turbulence statistics.

#### *3.2.1. Porosity*

The symmetry-breaking phenomenon is a result of the influence of geometric confinement on the pressure distribution on the solid obstacle surface. The porosity is the geometric parameter that controls the degree of flow confinement in the porous medium. The Reynolds number of the flow is fixed at $Re_p = 1{,}000$ and the porosity is decreased in the range $0.80 \geq \varphi \geq 0.43$ (Table 3). The degree of flow symmetry-breaking can be measured by using the macroscale flow angle $\theta_{macro}$, introduced in (3.1.1). The macroscale flow angle varies continuously with the porosity between zero before symmetry-breaking and a finite value after symmetry-breaking (Figure 14(a)). The flow transitions from symmetric to deviatory flow at the critical value of porosity, which lies between 0.67 and 0.72 for the case of the circular cylinder solid obstacles. Symmetry-breaking is dependent on two conditions that are related to the porosity – (1) asymmetry in the stagnation pressures before symmetry-breaking, and (2) the proximity of the solid obstacle surfaces. Together, these conditions create the lateral favorable pressure in the secondary flow region. This is why symmetry-breaking is non-existent at higher porosities even though the von Karman instability is present.

| Case ID | Porosity ($\varphi$) | Flow Properties |
|---|---|---|
| B1 | 0.43 | Deviatory flow – second configuration |
| B2 | 0.50 | Deviatory flow – second configuration |
| B3 | 0.56 | Deviatory flow – second configuration |
| B4 | 0.61 | Deviatory flow – second configuration |
| B5 | 0.67 | Deviatory flow – first configuration |
| B6 | 0.72 | Deviatory flow – first configuration |
| B7 | 0.80 | Symmetric flow |

Table 3: The LES cases simulated to analyze the dependence of symmetry-breaking on the porosity $\varphi$. The solid obstacles are circular cylinders and the Reynolds number ($Re_p$) is 1,000 for all of these cases.

When the porosity is less than the critical value for symmetry-breaking ($\varphi < 0.72$), two flow configurations are observed within the deviatory flow regime. The first is observed in the range $0.61 < \varphi < 0.72$, where there is flow separation on either side of the plane of geometric symmetry. There exists an asymmetry in the vortex core diameter of the micro-vortices formed on either side of the plane of geometric symmetry. The micro-vortices enter the primary flow region and they diminish in size in the converging-diverging section of the solid obstacle geometry. The second configuration is observed in the range $0.43 < \varphi < 0.61$, where the flow separation occurs on the same side of the plane of geometric symmetry. The degree of asymmetry in the vortex core diameter is less in the second configuration than in the first configuration. In the second configuration, the micro-vortices impinge on the solid obstacles in the lateral direction.

Deviatory flow is a source of macroscale turbulence anisotropy in homogeneous porous media with symmetric solid obstacles. The non-diagonal terms in the macroscale Reynolds Stress Tensor (RST) are typically zero in such porous media because the microscale RST is an even function about the plane of geometric symmetry (Chu *et al.* 2018). The microscale and macroscale RST are defined in equations (3.2) and (3.3), respectively. The variable $u'$ refers to the microscale velocity fluctuation.

$$\tau_{micro,ij} = \langle u_i{}' u_j{}' \rangle \tag{3.2}$$

$$\tau_{macro,ij} = \varphi \langle \langle u_i{}' u_j{}' \rangle \rangle^i \tag{3.3}$$

The double averaged governing equations contain other terms than the Macroscale RST that would also be influenced by deviatory flow. Terms such as the turbulent dispersion stress will also need to be considered during modeling. A comprehensive budget of the double averaged equations is beyond the scope of this paper but could be a valuable study in future work. In this paper, the analysis of the Macroscale RST is adequate to indicate the implications of symmetry-breaking on macroscale turbulence anisotropy. Since turbulence dispersion stress has a similar form to the Macroscale RST, the qualitative observations made below about the principal axis of the Macroscale RST can be expected for the turbulence dispersion stress tensor as well. When the microscale flow symmetry is broken, the microscale RST ceases to be an even function about the plane of geometric symmetry. After volume averaging, the macroscale RST has non-zero non-diagonal terms (Figure 14(b)). The magnitude of $\tau_{macro,12}$ increases when the porosity decreases in the case of the first flow configuration ($0.61 < \varphi < 0.72$). The magnitude of $\tau_{macro,12}$ decreases afterwards when the porosity decreases in the case of the second flow configuration ($0.43 < \varphi < 0.61$). The magnitude of the *xy*- microscale RST $\tau_{micro,12}$ is higher in the micro-vortex region. In the first flow configuration, the magnitude and the degree of asymmetry in the vortex core diameter is high, which results in the higher magnitude of $\tau_{macro,12}$ in the first flow configuration. The components of the macroscale RST $\tau_{macro,13}$ and $\tau_{macro,23}$ are virtually zero, since they are three orders of magnitude lesser than the diagonal components. They are not exactly equal to zero since the simulation has not been averaged for infinite time.

Symmetry-breaking in the microscale flow field causes symmetry-breaking between the macroscale velocity vector and the macroscale RST. The principal axes of the macroscale RST are rotated by an angle $\theta_{RST}$ from the Cartesian axes. The orientation of the principal axis of the macroscale RST with respect to the Cartesian axes is determined by computing its eigenvectors. The direction vector of the macroscale velocity is not identical to the orientation of the principal macroscale Reynolds Stresses $\theta_{RST}$ (Figure 14(a)). In other words, the macroscale RST is not oriented in the direction of the macroscale flow. This suggests that the macroscale turbulence anisotropy cannot be taken into account by axis transformation. The non-diagonal terms in the macroscale RST will need to be explicitly modeled.

To summarize, the microscale flow breaks symmetry at a critical value of porosity due to a lateral favorable pressure gradient that sustains macroscale deviatory flow. After symmetry-breaking, the macroscale velocity vector is oriented at an angle $\theta_{macro}$ with respect to the $x$-axis. The influence of symmetry-breaking extends beyond the macroscale velocity vector. If the coordinate axes are rotated by an angle of $\theta_{macro}$ such that the axes are aligned with the macroscale velocity vector, the macroscale RST is not oriented along the transformed axes. The principal axes of the macroscale RST are not aligned with either the macroscale velocity vector or the Cartesian axes (Figure 14(c)). This indicates a complete breakdown of symmetry in both the microscale and macroscale flow without the presence of a single plane of symmetry in the flow.

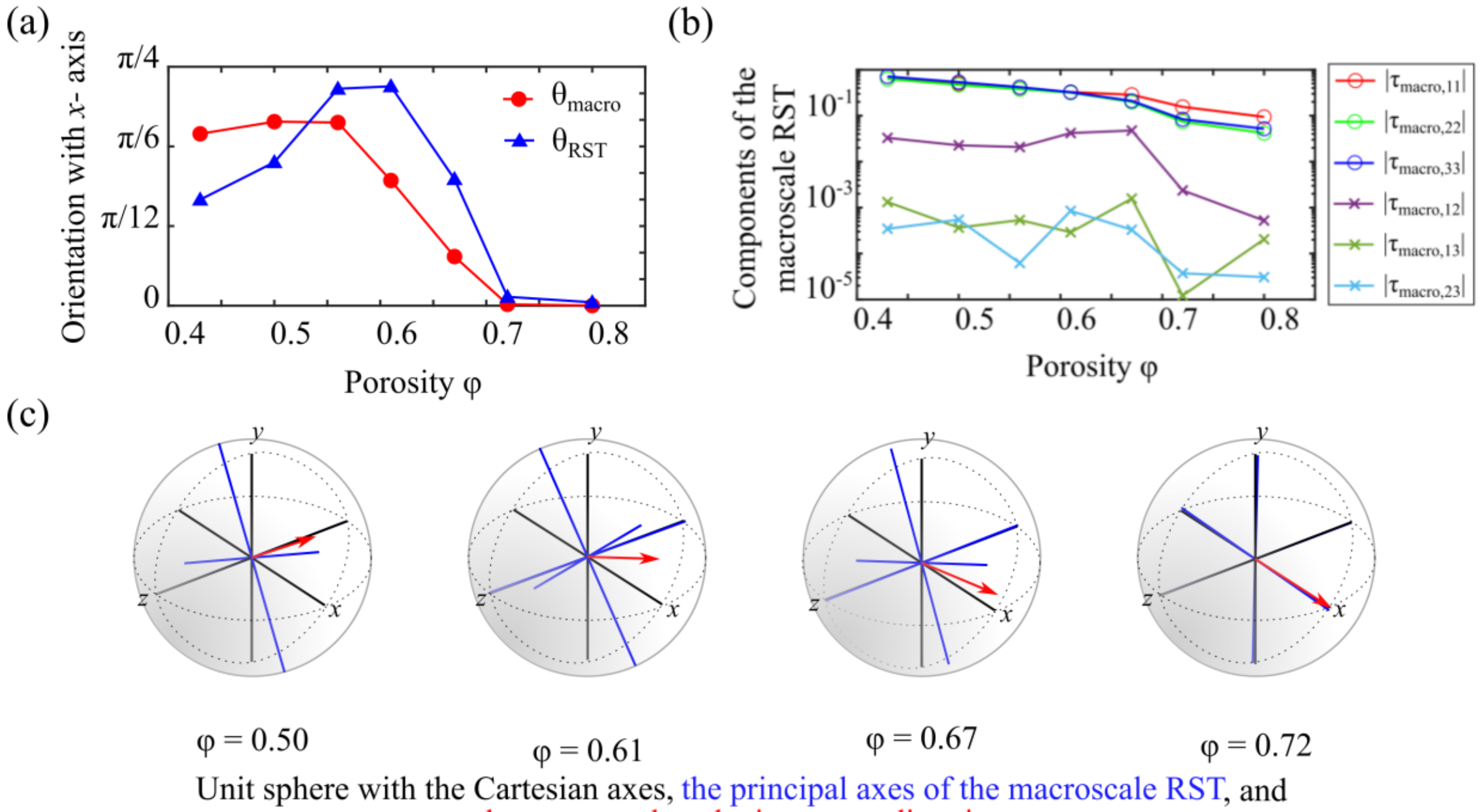

Figure 14: (a) The macroscale flow angle $\theta_{macro}$ and the principal axis of the macroscale RST projected on the $x$- axis $\theta_{RST}$, and (b) the components of the macroscale RST (note log scale on vertical axis), versus porosity $\varphi$ for $Re_p$ = 1,000. (c) The Cartesian axes (geometric axes), the principal axes of the macro RST, the macroscale velocity vector in a 3D unit sphere.

### *3.2.2. Reynolds number*

The Reynolds number is another critical parameter that determines the possibility of symmetry-breaking, since it controls the magnitude of the stagnation pressure in the microscale flow. The porosity is set equal to 0.5 and the Reynolds number is increased from $Re_p$ = 100 to 10,000 (Table 1). The critical value of Reynolds number for symmetry-breaking lies in between $Re_p$ = 300 to 500, and the details of its progression have been discussed in sections (3.1.1) and (3.1.2). A jump discontinuity is observed in the macroscale flow angle as the Reynolds number crosses the critical value. The critical Reynolds number is a necessary condition for the symmetry-breaking to occur. Unlike the porosity, the symmetry-breaking cannot be parameterized using the Reynolds number to control the deviatory flow. When the Reynolds number is changed, deviatory flow can either emerge or disappear resulting in a binary change with zero or a constant value of $\theta_{macro}$ depending on the porosity. Whereas, an increase in the porosity can result in a non-binary, incremental change in $\theta_{macro}$.

The anisotropy in the macroscale RST of turbulence after symmetry-breaking is evident in all values of Reynolds numbers above the critical value (Figure 15). Microscale turbulence is expected to become more isotropic as the Reynolds number increases (Wood *et al.* 2020). However, the deviatory flow after symmetry-breaking leads to an increase in $\tau_{macro,12}$ with the Reynolds number, which results in an increase in the macroscale anisotropy of turbulence. The consistent increase in $\tau_{macro,12}$ with the Reynolds number implies that the macroscale anisotropy is not formed from the microscale turbulence structures. In section (3.1.2), the microscale turbulence structures after symmetry-breaking were observed to be 3-D fine-scale structures as opposed to the 2-D large-scale structures observed before symmetry-breaking. Therefore, the origin of macroscale anisotropy is purely the asymmetry in the microscale flow topology arising from symmetry-breaking. Here, microscale flow topology refers to the microscale velocity and pressure distribution, the location of the primary and secondary flow regions, and the location and the number of flow separation and stagnation points.

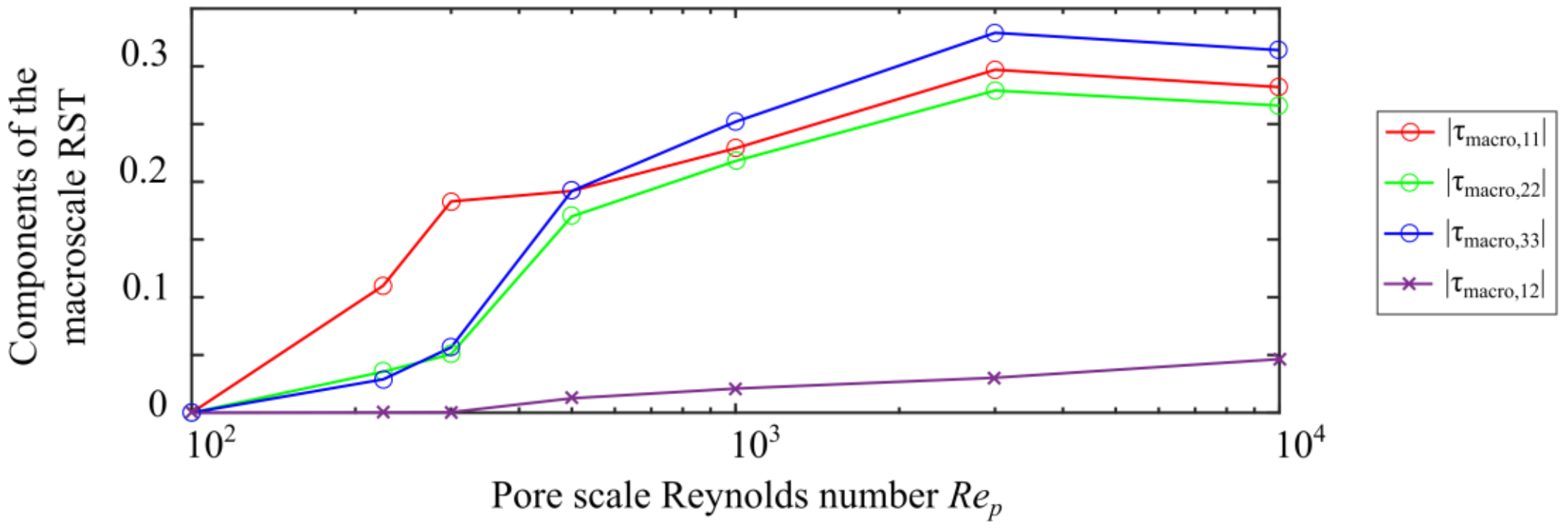

Figure 15: The components of the macroscale Reynolds stress tensor versus Reynolds number for $\varphi = 0.5$.

### *3.2.3. Solid obstacle shape*

The shape of the solid obstacles influences the vortex shedding process and the location of flow stagnation points, which are the source of the symmetry-breaking phenomenon. Therefore, deviatory flow cannot occur for all solid obstacle shapes. Take for instance a square cylinder solid obstacle forming a porous medium with a porosity of 0.5. This value of porosity is within the critical porosity for symmetry-breaking in a porous matrix with circular cylinder obstacles. Deviatory flow is not observed for the square cylinder solid obstacles for any of the simulated cases (Table 4). The Hopf bifurcation has been reported to occur for square solid obstacles (Agnaou *et al.* 2016). An asymmetric vortex pair is formed behind each solid obstacle (Figure 16). Microscale flow symmetry is broken as a result of these instabilities, but it does not get translated into a symmetry-breaking in the Reynolds-averaged macroscale flow field.

In the case of a square cylinder porous medium with low porosity (case C1), the flow separation and stagnation occur at the vertices of the square shape (Figure 16(b)). Thus, the locations of the flow separation and stagnation are predetermined by the geometry. A low value of porosity is required for the formation of a lateral favorable pressure gradient. The micro-vortices formed in this case are characterized by slow turn-over and resemble the flow in an open cavity flow (Figure 16(b)). The primary and secondary flow regions are distinct. The primary and secondary flow regions are separated by a strong shear layer (Figure 16(a)), which “channels” the primary flow with little intervention from the secondary flow. The secondary flow region is confined in-between the solid obstacles by the primary flow. A sufficient magnitude of lateral favorable pressure gradient is not observed when the porosity is increased (case C2) or when the Reynolds number is increased (case C3). The solid obstacle shape must possess a greater degree of circularity for the symmetry-breaking phenomenon to occur.

| Case ID | Porosity ($\varphi$) | Reynolds number ($Re_p$) |
|---|---|---|
| C1 | 0.50 | 1,000 |
| C2 | 0.67 | 1,000 |
| C3 | 0.50 | 10,000 |

Table 4: The LES cases simulated to analyze the dependence of symmetry-breaking on the solid obstacle shape. The solid obstacle shape is square cylinder for all of these cases.

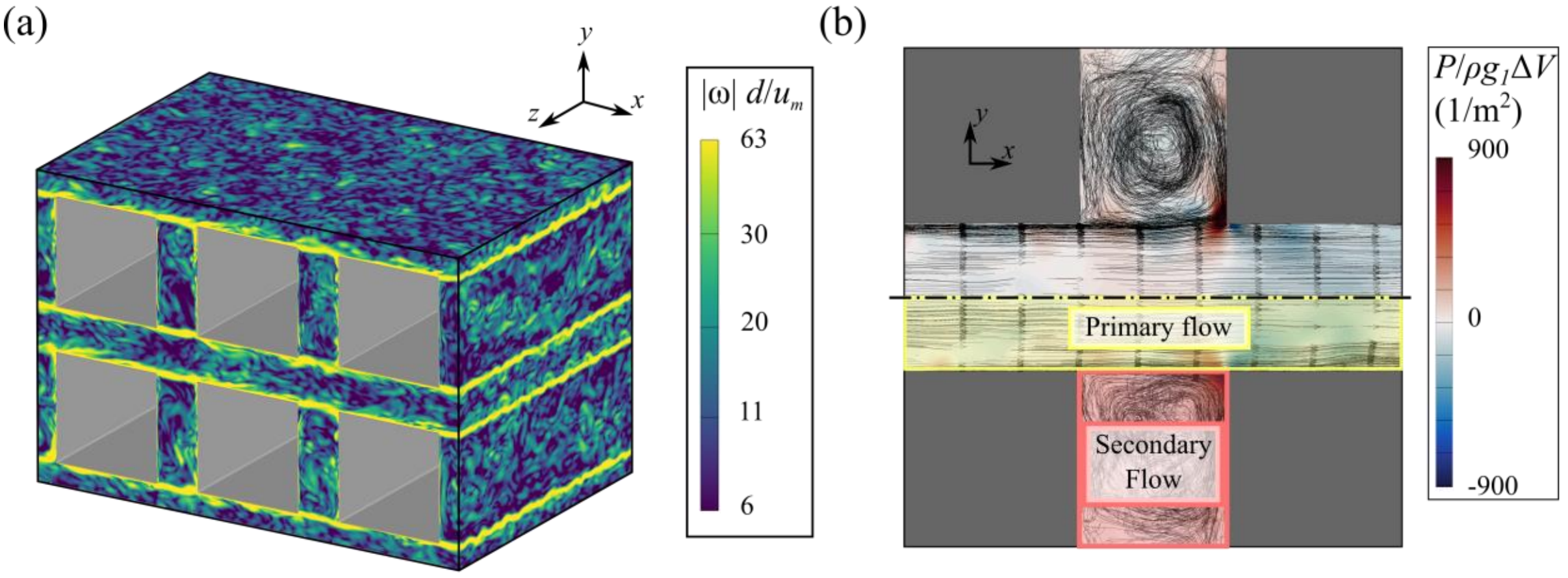

Figure 16: (a) Turbulent structures visualized using the 2D contours of vorticity magnitude for a porous medium with square cylinder obstacles and porosity $\varphi = 0.5$ (case C1). A sub volume of dimensions (3*s*, 2*s*, 2*s*) of the REV-T is shown in this figure. (b) Instantaneous flow streamlines projected on the *xy*- plane for a sub volume of dimensions (*s*, *s*, 2*s*). The streamlines are plotted on top of contours of static pressure.

## 4. Summary

A symmetry-breaking phenomenon is reported to occur in microscale turbulent flow inside periodic porous media with a low porosity. The symmetry-breaking phenomenon results in a macroscale deviatory flow that is oriented with the direction of applied pressure gradient by an angle $\theta_{macro}$. The Reynolds-averaged deviatory flow satisfies the conservation of momentum in the lateral direction since the viscous and the pressure components of the lateral drag force have equal magnitude and act in opposite directions. In the instantaneous flow field, symmetry-breaking originates from a low-frequency flow instability that is formed by the competition of the macroscale inertial and pressure forces. The formation of the low-frequency flow instability results in high-amplitude oscillations of the pressure force. High-frequency oscillations of the pressure force due to the von Karman instability in vortex shedding are superimposed on the low-frequency oscillations. The transition to turbulence precedes the occurrence of the symmetry-breaking phenomenon. Turbulence in the flow aids the symmetry-breaking phenomenon through the breakup of micro-vortices and the increase in the adverse pressure gradient. It also leads to the formation of random modes of deviatory flow. The occurrence of the symmetry-breaking phenomenon is marked by the disappearance of the low- and high-frequency oscillations from the power spectrum. Instead, the power spectrum consists of a band of frequencies that are characteristic of turbulence. When the applied pressure gradient is increased such that the Reynolds number is above the critical value for symmetry-breaking, the amplitude of the oscillation of the pressure forces increases to a critical point. The high amplitude of the pressure force induces symmetry-breaking and flow in the lateral direction. After symmetry-breaking, the amplitude of oscillation of the pressure force reduces. This shows that symmetry-breaking diminishes the highly unsteady flow instabilities that were present before symmetry-breaking. The phase difference that is present in the pressure forces acting on the individual solid obstacles leads to the formation of the modes of deviatory flow. In infinitely periodic porous media, infinite combinations of the two directions of symmetry-breaking are possible behind each solid obstacle. Each combination will result in a unique mode of the deviatory flow. The direction of symmetry-breaking is influenced by the phase of vortex shedding behind each solid obstacle and its immediate neighbors.

Symmetry-breaking first emerges in the microscale flow field, which is then transferred to the macroscale flow through the pressure drag force. The microscale flow field before symmetry-breaking consists of recirculating vortex tubes that deform to form a knit pattern of 3D reverse hairpin vortices. The micro-vortices form flow stagnation points on the solid obstacle surface on either side of the plane of geometric symmetry. The stagnation points are located in the converging portion of the GPM geometry. A net adverse pressure gradient is experienced in the direction of applied pressure gradient as a result of the flow stagnation. When the applied pressure gradient is changed to induce symmetry-breaking, the flow structures break down from the orderly knit pattern of large hairpin structures to random fine-scale structures. Thus, the recirculating vortex tube is broken down into a recirculation zone consisting of smaller turbulent structures. The adverse pressure gradient experienced by the flow in the primary flow region breaks up the large-scale turbulent structures into fine-scale turbulent structures. Meanwhile, a non-linear increase in the magnitude of flow stagnation pressure increases the strength of the adverse pressure gradient, forming a pressure plug in the converging portion of the GPM. The breakup of a recirculation vortex system triggers microscale symmetry-breaking through a lateral favorable pressure gradient. Symmetry-breaking propagates away from the location of incidence resulting in deviatory flow throughout the REV. The direction of deviatory flow and the formation of the modes are decided by the phase of vortex shedding at the time of symmetry-breaking. Deviatory flow is sustained by the lateral favorable pressure gradient provided the Reynolds number is maintained above the critical value.

Deviatory flow due to symmetry-breaking alters the macroscale properties of turbulence in porous media. The asymmetry in the microscale flow field results in non-zero non-diagonal components of the macroscale Reynolds Stress Tensor (RST), especially in the plane of symmetry-breaking. The principal axes of the macroscale RST form a 3D angle with the geometric axes. The orientation of the principal axes does not coincide with either the macroscale velocity vector or the geometric axes, signifying a complete breakdown of symmetry in the flow. Therefore, flow symmetry-breaking is a source of macroscale turbulence anisotropy in symmetric porous media.

The occurrence of symmetry-breaking is dependent on the porosity, Reynolds number, and solid obstacle shape. Critical values of porosity and Reynolds number are required as a necessary condition for symmetry-breaking. The symmetry-breaking phenomenon is driven by the presence of a high magnitude of lateral favorable pressure gradient in a confined geometry. Two configurations of microscale flow separation about the plane of geometric symmetry occur based on the degree of confinement in the geometry. They possess unique properties due to the difference in the location of flow separation and stagnation. The circularity of the solid obstacle shape also decides the location of flow separation and stagnation. Therefore, the circularity of the solid obstacle shape is a necessary condition for symmetry-breaking as well.

**Acknowledgment**

AVK acknowledges the support of the National Science Foundation (award CBET-1642262), and the Alexander von Humboldt Foundation through the Humboldt Research Award. The authors wish to thank Haodong Li for helping to prepare figure S3 in the supplementary material.

## Appendix A. Validation of the size of the REV-T

In this section, the convergence of the macroscale flow solution at the chosen REV size is demonstrated. Case B7 from Table 3 is chosen as the representative case. The turbulent structures will be restricted the least at the porosity of 0.8 by the solid boundaries of the obstacles. If the REV size is adequate for $\varphi = 0.8$, it is reasonable to assume that it will be adequate for cases with lower porosity. At $\varphi = 0.8$, the solid obstacle shape has the least influence on the macroscale properties among all of the cases. A Reynolds number of 1,000 is chosen since a majority of cases assume this value or lesser. Unlike $Re_p = 225$, the flow at $Re_p = 1{,}000$ is not close to the turbulence transition point and is, therefore, representative of the limiting cases of turbulence transport in this paper with regard to REV size.

With these parameters fixed, the REV size is increased from $1s$ to $5s$ in increments of $1s$. A grid resolution of $0.02s$ is used to perform LES as per section (2.2.1) for each of the REV sizes. Two macroscale quantities are analyzed in this study: the mean applied momentum source $\langle g_i \rangle$, and the macroscale TKE $\langle k \rangle^i$ (Figure 17). The applied momentum source term is directly linked to the drag force on the surface of the solid obstacles inside the porous medium. The macroscale TKE indicates whether an adequate sample of the turbulence fluctuations is captured by the REV. Both $\langle g_i \rangle$ and $\langle k \rangle^i$ converge at an REV size of $4s$ with changes of only 0.4% and 0.25% respectively when the REV size is increased to $5s$. When the size of the REV is increased from $1s$, there is a staggered trend observed depending on whether an odd or even number of obstacles are present in the REV. The cause of the staggering is a decoupling between the odd and even number REVs. The decoupling is brought about by the influence of the periodic boundary condition that is imposed, and also the number of modes of the microscale flow instability that can be present in the domain. The distinction between the odd and even number REVs diminishes at the REV size of $4s$. This offers further confirmation that the REV size of $4s$ is adequate for the simulations presented in this paper. We decreased the REV size in the $z$-direction for cylindrical solid obstacle geometries from $4s$ to $2s$. The turbulence two-point correlation function was found to de-correlate in the $z$- direction in a span of $1s$. The turbulent structure visualizations also showed smaller structure size in the $z$- direction.

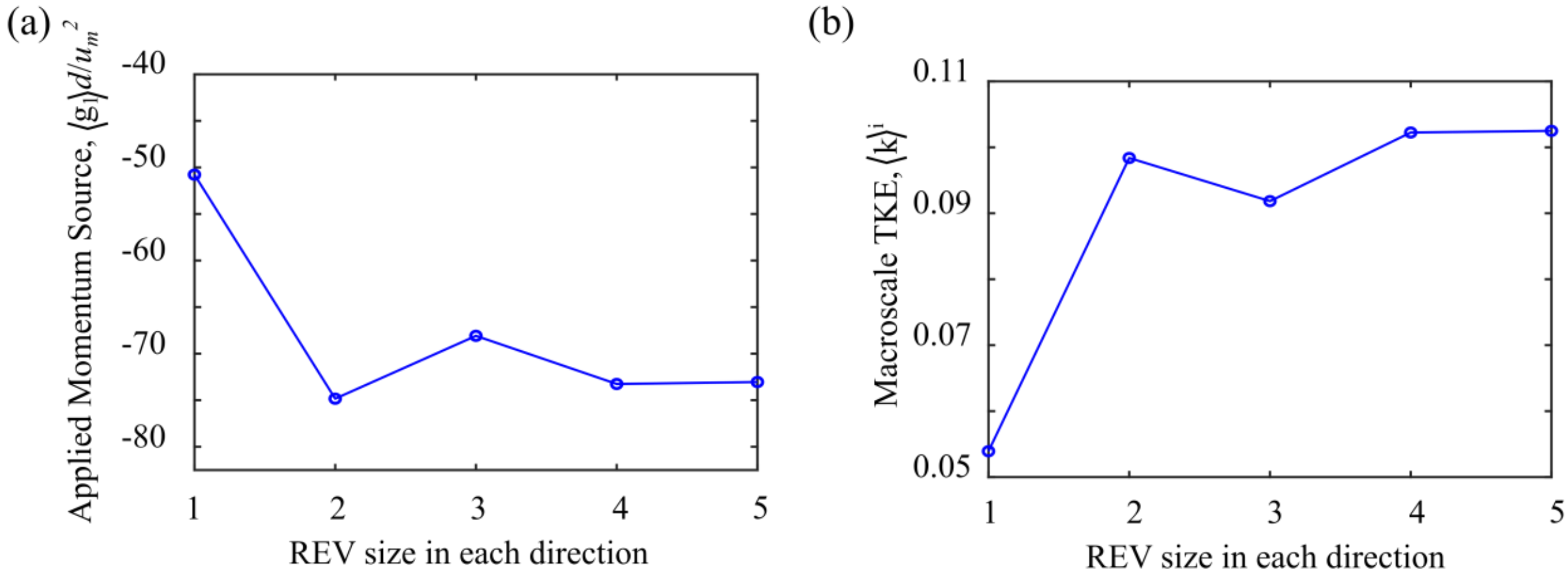

Figure 17: The convergence of the (a) applied momentum source and (b) the macroscale TKE with increase in the REV size.

## Appendix B. Validation of the numerical grid

In this section, the adequacy of the grid resolution is established for the LES and DNS methods. The grids are designed using an *a priori* estimation using the Taylor microscale $\lambda$. A majority of simulations that are presented in this work adopt a Reynolds number of 1,000 or less. The value of $\lambda/d$ at this Reynolds number is estimated as 0.1 according to the definition given in Pope (2000). The porosity of the medium is varied from 0.80 to 0.43. Since the size of the largest eddy is expected to be of the size of the pore, grid resolution tests are performed for an REV with a single solid obstacle by parameterizing the cell sizes (non-dimensionalized by the pore size). The grid resolution tests have been performed for 4 values of porosity, $\varphi$ = 0.50, 0.61, 0.72, and 0.80, of a medium with circular cylinder solid obstacles (see Table 5 and Table 6). A Reynolds number of 1,000 has been chosen for the tests. The idea behind performing the grid resolution test is to determine whether the contribution of the energetic scales of motion is captured.

For the Scale-Resolving Schemes (LES and DNS), the turbulence kinetic energy spectrum is used to identify the scale regimes of turbulence that have been resolved in this work. The turbulence kinetic energy of the inertial range of eddies is characterized by a (-5/3) slope in the wavenumber domain according to the Kolmogorov Similarity Hypothesis. The velocity fluctuation correlation function $R_{ij}$ is calculated at the midpoint of the void volume using equation A 1. The operator $\langle - \rangle$ denotes Reynolds averaging, $x_0$ is located at the centroid of the void space (see Figure 1(a)), and $r$ varies along the $z$- direction. The Fourier transform of the correlation function $R_{ij}$ is obtained using an FFT routine and the energy spectrum is computed as the Power Spectral Density of the Fourier-transformed correlation function. The present formulation of the turbulence energy spectrum assumes isotropy. It is used only to serve as a measure of grid resolution. The turbulence kinetic energy spectra versus the wavenumber ($k \cdot s$) for the LES test cases are shown in Figure 18.

$$R_{ij}(r, x_0) = \langle u'_i(x_0) u'_j(x_0 + r) \rangle \quad \text{(A 1)}$$

| Porosity φ | Coarse grid, $\Delta x_{max}/s$= 0.03 | Intermediate grid, $\Delta x_{max}/s$= 0.02 | Fine grid, $\Delta x_{max}/s$= 0.01 |
|---|---|---|---|
| 0.50 | 0.65 | 0.76 | 0.76 |
| 0.61 | 0.55 | 0.56 | 0.72 |
| 0.72 | 0.48 | 0.53 | 0.58 |
| 0.80 | 0.53 | 0.42 | 0.49 |

Table 5: The maximum value of non-dimensional near-wall grid spacing, $\Delta y^{+}_{max}$, measured on the surface of the solid obstacles for the grid resolution test cases. The value of the grid size $\Delta x_{wall}/s$ was set equal to 0.001 for all of the cases.

The energy spectrum plots show that the dissipative scales of turbulence are not resolved by these grids. A grid resolution of $\Delta x_{max}/s = 0.03$ is not sufficient to capture the inertial subrange. The grids with $\Delta x_{max}/s = 0.02$ offer a reasonable compromise between grid resolution and computation time for LES. It is sufficient to capture the large-scale turbulent vortex system that scales with $(s - d)$, which corresponds to a ratio of 10:1 with $s$. The turbulence energy spectra for all three grid sizes are almost identical for $k \cdot s$ = 1-10 with only the coarse grid starting to show deviation at $k \cdot s = 10$. The LES cases in subsequent sections will adopt $\Delta x_{max}/s$ = 0.02 as the grid resolution. The DOETKE subgrid model (section 2.2.1) was set active to capture the unresolved scales of turbulence. A quality measure known as the LES Index of Quality or LES_IQ (Celik *et al.* 2005) has also been calculated for the LES test cases. LES_IQ provides the fraction of the total turbulence kinetic energy that is resolved by the grid. Pope (2004) recommends that 80% of the energy must be resolved in LES, resulting in a quality criterion of LES_IQ > 0.8. Celik *et al.* (2005) remark that simulations with LES_IQ > 0.9 may be considered to be of DNS quality. This is the case for all of the LES simulations in this work after time averaging. The minimum and spatially averaged values of LES_IQ at an instant in time are reported in Table 6. The minimum instantaneous value is less than 0.8 for all of the test cases. The minima are located at the core of the small streaks of turbulent structures that interact with the solid obstacles. The streaks are local, and the low LES_IQ that is associated with them will vanish when it is averaged in the $z$- direction. Premature dissipation of eddies in the near-wall region is expected in this work, which is derived from the diffusive nature of the subgrid model. For the DNS simulation, a grid size of $\Delta x_{max}/s = 0.0095$ is chosen. In Figure 18, the energy density for $\Delta x_{max}/s = 0.01$ at $k \cdot s = 100$ is smaller than that of the largest eddies by a factor of $10^{-4}$ for all the tested values of porosity. Further grid refinement will result in a change in the solution that is not paramount to the study of the large-scale motions. The primary reason for DNS resolution is to minimize the error, which is less than 10% throughout the DNS case.

| Porosity φ | | Coarse grid, $\Delta x_{max}/s$= 0.03 | Intermediate grid, $\Delta x_{max}/s$= 0.02 | Fine grid, $\Delta x_{max}/s$= 0.01 |
|---|---|---|---|---|
| 0.50 | minimum | 0.37 | 0.42 | 0.54 |
| | average | 0.90 | 0.94 | 0.98 |
| 0.61 | minimum | 0.56 | 0.63 | 0.66 |
| | average | 0.93 | 0.95 | 0.98 |
| 0.72 | minimum | 0.48 | 0.48 | 0.49 |
| | average | 0.93 | 0.96 | 0.99 |
| 0.80 | minimum | 0.46 | 0.46 | 0.60 |
| | average | 0.93 | 0.96 | 0.99 |

Table 6: The value of LES_IQ measured in the fluid volume for the grid resolution test cases. Both the minimum and the volume-averaged values are reported (ranges from 0 to 1, high values indicate high resolution with a large fraction of the turbulence kinetic energy being resolved).

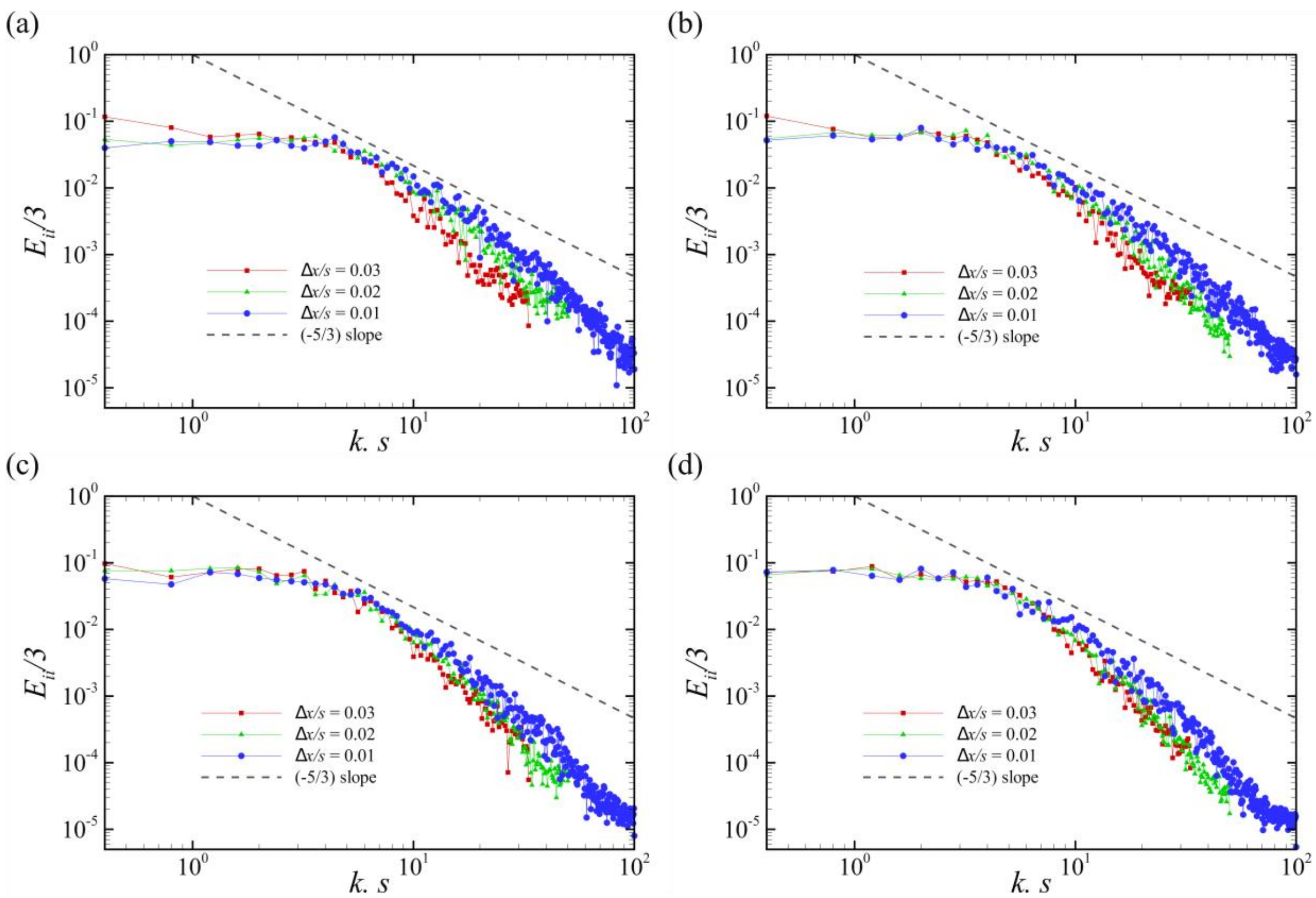

Figure 18: Turbulence energy spectrum for LES cases (a) φ = 0.50, (b) φ = 0.61, (c) φ = 0.72, (d) φ = 0.80 at $Re_p$ = 1,000. The red square data points correspond to $\Delta x_{max}/s$ = 0.03, green triangles to $\Delta x_{max}/s$ = 0.02, blue circles to $\Delta x_{max}/s$ = 0.01, the dashed line to the -5/3 slope on the log plot.

For DNS, the grid study is performed by repeating the LES cases with a high order DNS solver, Incompact3d (Laizet & Lamballais 2009). The purpose of using Incompact3d is to independently verify the occurrence of the symmetry-breaking phenomenon and to confirm the ability of ANSYS Fluent to perform DNS calculations. For the Incompact3d simulations, the Finite Difference Method (FDM) is used with 6$^{th}$ order spatial discretization and 2$^{nd}$ order explicit time integration. The spectral method is used to solve the pressure Poisson equation. Cartesian grids are used for the REV and the solid obstacles are imposed using the Immersed Boundary Method. The symmetry-breaking phenomenon is observed in the DNS simulations using Incompact3d. The resulting turbulence energy spectra are plotted in Figure 19. The magnitude range of the turbulence kinetic energy is similar for the LES and DNS cases. For the wavenumber range $k \cdot s$ = 10-100, the features of the turbulence energy spectrum at the different values of porosity are also similar. For φ = 0.5 and 0.61, the spectra in $k \cdot s$ = 10-100 for the three grid resolutions are spaced apart. For φ = 0.72 and 0.8, the spectra in $k \cdot s$ = 10-100 are identical. Therefore, the observations made for the LES simulations can be extended to the DNS simulations. For the DNS simulations, a grid resolution of $\Delta x_{max}/s$ = 0.0095 is finer than any of the cases shown in Figure 18. ANSYS Fluent is used for DNS since it supports body-fit grids.

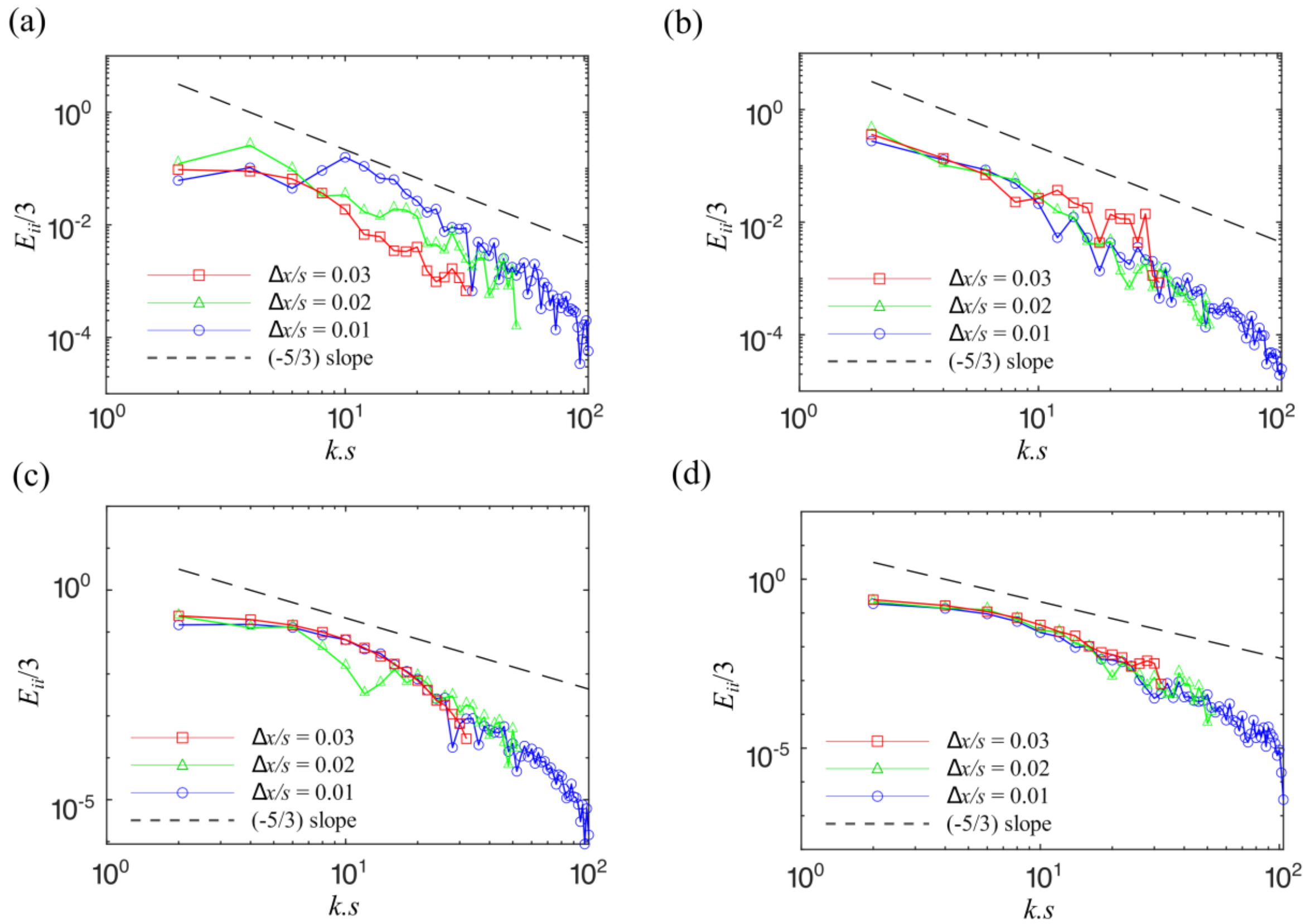

Figure 19: Turbulence energy spectrum for DNS cases with Incompact3d (a) φ = 0.50, (b) φ = 0.61, (c) φ = 0.72, (d) φ = 0.80 at $Re_p$ = 1,000. The red square data points correspond to $\Delta x_{max}/s$ = 0.03, green triangles to $\Delta x_{max}/s$ = 0.02, blue circles to $\Delta x_{max}/s$ = 0.01, the dashed line to the -5/3 slope on the log plot.

**Appendix C. Surface stress distribution for the Reynolds averaged flow in case A4**

In this appendix, we provide additional plots that support Figure 2(a). The momentum conservation in the *y*- direction for the deviatory flow for the Reynolds averaged flow field is ensured by the *y*- direction pressure and viscous forces (defined in equation 3.1). The inertial and applied forces are zero in the *y*- direction for the Reynolds averaged flow. The pressure and viscous stresses are distributed on the solid obstacle surface such that the *y*- direction resultant force is zero. Figure 20 shows the *y*- direction pressure and viscous stress distributions on the surface of a single solid obstacle for case B2 as per the definition given under the integral symbol in equation 3.1. Note that the deviatory flow results in asymmetric stress distributions on the surface of the solid obstacle. The resulting pressure and viscous forces in the *y*- direction are non-zero in deviatory flow (unlike in symmetric flow). The forces are calculated by integrating the stress distributions over the solid obstacle surface as given in equation 3.1. Therefore, it is important to note that a point-by-point match between the pressure and viscous stresses is not expected.

The deviatory flow forms a high stagnation pressure region in the lower-left quadrant of the cylindrical solid obstacle. Therefore, the *y*- pressure force is going to act in a positive direction. In case B2, the *y*- pressure force per unit volume non-dimensionalized by $\rho g_1 d$ is 2.7931. The *y*- direction force magnitude is 5% of that in the *x*- direction. Note that we do not apply any pressure gradient in the *y*- direction, but we apply a pressure gradient in the *x*- direction to sustain the flow. Therefore, we consider the *y*- direction forces to be a significant contribution brought by symmetry-breaking. The resultant instantaneous pressure force in the *y*- direction causes the deviatory flow. The micro-vortices are located on the top-right quadrant of the cylinder resulting in low shear magnitudes in that location. The bifurcation of the flow at the stagnation point (lower-left quadrant) forms high magnitudes of shear stress. Due to a lack of symmetry in the shear stress distribution, there is a net *y*- shear force in the negative direction. The *y*- viscous force per unit volume non-dimensionalized by $\rho g_1 d$ is -2.8074. Since momentum must be conserved, the *y*- pressure and the *y*- shear forces have equal magnitudes so that they cancel each other by acting in opposing directions. The numerical error of 0.0143 is 4 orders of magnitude less than the magnitude of the applied pressure gradient. The small numerical error arises from the finite time interval used for sampling the flow variables. While 4 orders of magnitude is an adequate tolerance for the statistically averaged variables in this study, note that the error will decrease further the closer we are able to get to an "infinite" sampling time interval.

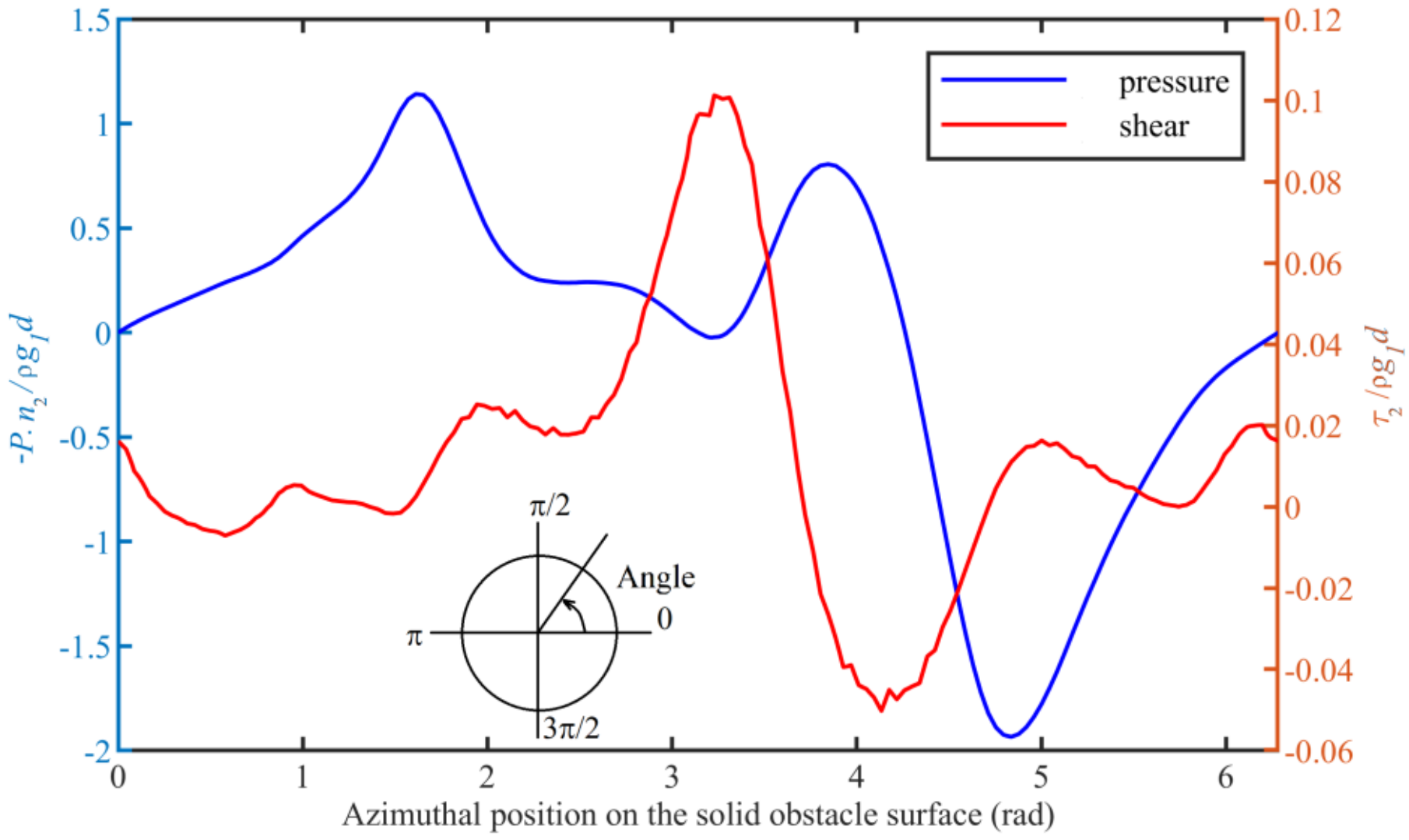

Figure 20: The distribution of the *y*- direction pressure -$P \cdot n_2$ (blue) and shear $\tau_2$ (red) stresses on the solid obstacle surface for case B2.